\documentclass[11pt]{article}

\usepackage{pgfplots}
\usepackage{pgfkeys}
\usepackage{amsmath}
\usepackage{amsfonts}
\usepackage{amssymb}
\usepackage{color,graphicx}
\usepackage{epstopdf}
\usepackage{graphics}
\usepackage{subfigure}
\usepackage{amsthm}
\usepackage{pdfsync}

\usepackage{hyperref}
\usepackage{fullpage}

\newcommand{\inc}{\textrm{inc}}

\usepackage{MesCommandes}

\usepackage{mathrsfs}        
\usepackage{amsmath}
\usepackage{amsfonts}
\usepackage{amssymb}
\usepackage{stmaryrd}        
\usepackage{array}           
\usepackage{epsfig}

\usepackage{color}
\usepackage{amsthm}
\usepackage{amsmath}
\usepackage{amsfonts}
\usepackage{amssymb}
\usepackage{graphicx}
\usepackage{graphics}
\usepackage{subfigure}

\usepackage{import} 

\usepackage{pdfsync}

\usepackage{amsthm}          

\theoremstyle{plain} 

\newtheorem{prop}{Proposition}  

{\everymath{\displaystyle\everymath{}}\array}%
{\endarray}

\makeatletter
\newenvironment{myalign*}{%
  \@fleqntrue\csname align*\endcsname
  }{%
  \csname endalign*\endcsname
  } 

\makeatletter
\let\l@theoreme\l@figure
\makeatother
\makeatletter
\newcommand{\listoftheorems}{%
\chapter*{Liste des théorèmes et propositions}\@starttoc{lthm}}
\makeatother

\usepackage{amsopn}          

\newcommand{\kint}{k^-}
\newcommand{\kintp}{\kint_p}
\newcommand{\Nscat}{M}
\newcommand{\ups}{u^+}
\newcommand{\um}{u^-}
\newcommand{\rhops}{\rho^+}
\newcommand{\rhomi}{\rho^-}

\usepackage{xcolor}

\title{$\mu$-diff: an open-source Matlab toolbox for computing multiple scattering problems by disks}

\author{Bertrand Thierry\footnotemark[1] , Xavier Antoine\footnotemark[2] , Chokri Chniti\footnotemark[3] \    and 
Hasan Alzubaidi\footnotemark[3]}

\date{}

\begin{document}

\maketitle

\renewcommand{\thefootnote}{\fnsymbol{footnote}}

\footnotetext[1]{Laboratoire J.L. Lions (LJLL), University of Paris VI, Paris, France. ({\tt bthierry@math.cnrs.fr}).}
\footnotetext[2]{Universit\'e de Lorraine, Institut Elie Cartan de Lorraine, UMR 7502, Vandoeuvre-l\`es-Nancy, F-54506, France. ({\tt xavier.antoine@univ-lorraine.fr}).}
\footnotetext[3]{Department of Mathematics, University College in Qunfudah, Umm Al-Qura University, Saudi Arabia.
 ({\tt cachniti@uqu.edu.sa;hmzubaidi@uqu.edu.sa}).}

\renewcommand{\thefootnote}{\arabic{footnote}}

\begin{abstract}
The aim of this paper is to describe a Matlab toolbox, called $\mu$-diff, 
 for modeling and numerically solving two-dimensional complex multiple scattering by a large
collection of circular cylinders. The approximation methods in $\mu$-diff are based on the Fourier series expansions of 
the four basic integral operators arising in scattering theory. Based on these expressions,
an efficient spectrally accurate finite-dimensional solution of multiple scattering problems can be simply obtained
for complex media even when many scatterers are considered as well as large frequencies.
The solution of the global linear system to solve can use either direct solvers or preconditioned 
iterative Krylov subspace solvers for block Toeplitz matrices. Based on this approach, this paper
explains how the code is built and organized. Some complete numerical examples of applications (direct and inverse scattering)
are provided to show that $\mu$-diff is a flexible, efficient and robust toolbox for solving some complex multiple scattering problems.
\end{abstract}

\medskip




\medskip




\tableofcontents

\section{Program Summary}

\textit{
\hspace{-1.8em}Manuscript title: $\mu$-diff: an open Matlab toolbox for computing multiple scattering problems by disks
\\ Authors: Xavier ANTOINE \& Bertrand THIERRY
\\ Program title: $\mu$-diff
\\ Licensing provisions: Standard CPC licence
\\ Programming language: Matlab
\\ Computer(s) for which the program has been designed: PC, Mac
\\ Operating system(s) for which the program has been designed: Windows, Mac OS, Linux
\\ RAM required to execute with typical data: 2000 Megabytes
\\ Has the code been vectorised or parallelized?: Yes
\\ Number of processors used: Most if not all
\\ Keywords: Matlab, Multiple scattering, waves, random media, acoustics, optics, electromagnetism, numerical methods
\\ CPC Library Classification:  4.6, 10, 18
\\ Nature of problem: Modeling and simulation of two-dimensional multiple wave scattering by large clusters of circular cylinders for any frequency.
The program is well-designed to manage highly accurate solutions for deterministic or random media, with various boundary conditions 
and physics properties of the scatterers. Pre- and post-processing facilities are designed specifically for these problems.
\\ Solution method: We use spectral Fourier approximation schemes and direct or iterative Krylov subspace methods.
\\ Running time: From a few seconds for simple problems to a few minutes for more complex situations on a medium computer.
}

\section{Introduction}

Let us consider $M$ regular, bounded and disjoint scatterers $\Omega^-_p$, $p=1,...,M$, distributed in 
$\mathbb{R}^2$,  with boundary  $\Gamma_p:=\Omega_p^-$. The scatterer
 $\Omega^-$ is defined as the collection of the $M$ separate obstacles, i.e. $ \Omega^- = \cup_{p=1}^M \Omega^-_p$, with boundary
  $ \Gamma = \cup_{p=1}^M \Gamma_p$.
 The homogeneous and isotropic exterior domain of propagation is $\Omega^+=\mathbb{R}^2\setminus\overline{\Omega^-}$.
 For the sake of conciseness in the presentation, we first assume that the scatterers are sound-soft (Dirichlet boundary condition), but other 
situations can be handled by the $\mu$-diff (\textbf{mu}ltiple-\textbf{diff}raction) Matlab toolbox (e.g. sound-hard scatterers, impedance boundary conditions, 
penetrable scatterers)
as it will be shown during the numerical examples (see section \ref{sectionnumericalexamples}).
We now consider a time-harmonic incident  acoustic plane wave  $u^{\inc}(\mathbf{x}) = e^{ik \Beta \cdot \mathbf{x}}$
(with $\mathbf{x}=(x_1, x_2) \in \mathbb{R}^2$) illuminating $\Omega^-$, with an incidence direction  $\Beta = 
(\cos(\beta), \sin(\beta))$ and a time
 dependence  $e^{-i\omega t}$, where $\omega$ is the wave pulsation and $k$ is 
the wavenumber.
The sound-soft multiple scattering problem  of $u^{\textrm{inc}}$ by $\Omega^-$ consists in computing
 the scattered wavefield $u$ as the solution to the boundary-value problem \cite{AntGeuRam10,Mar06}
\begin{equation}
\label{eqEqInt:ProblemeU}
\left\{\begin{array}{r c l l}
\displaystyle  (\Delta + k^2) u &=& 0, &\textrm{ in $\Omega^+$,} \\
\displaystyle  u &=& -u^{\inc},  &\textrm{ on $\Gamma$,} \\
\multicolumn{4}{l}{\displaystyle {\lim_{||\mathbf{x}||\to + \infty} ||\mathbf{x}||^{1/2}\left(\nabla u\cdot \frac{\mathbf{x}}{||\mathbf{x}||}-i k u\right)=0.} }
\end{array}\right.
\end{equation}
The  operator  $\Delta = \partial_{x_1}^2 + \partial_{x_2}^2$ is the Laplace operator
and $(\Delta + k^2)$ is the Helmholtz operator. The  gradient operator is  $\nabla$ and
$||\mathbf{x}||=\sqrt{\mathbf{x} \cdot \mathbf{x}}$, where $\mathbf{x}\cdot \mathbf{y}$ is the scalar product of two vectors
$\mathbf{x}$ and $\mathbf{y}$ of $\mathbb{R}^2$. The last equation of  (\ref{eqEqInt:ProblemeU})
is the well-known Sommerfeld's radiation condition at infinity that ensures the uniqueness of  $u$ \cite{ColKre83,Ned01}.

Multiple scattering is known to be a very complicated and challenging problem in terms of computational method 
\cite{AcostaOSRC,AcostaVillamizar,AntChnRam08,AntGeuRam10,JACT,ChenLeeLin,MEhrhardtBook,MHZ,GeuzaineBrunoReitich,Grote2004630,KharlamovFilip,Van} since the incident wave is multiply diffracted by all the single scatterers involved in the geometrical configuration.
As a consequence, the scattered wavefield has a highly complicated structure and exhibits some particular physics properties.
The toolbox $\mu$-diff\footnote{\url{http://mu-diff.math.cnrs.fr}} contributes to the development of  reliable and efficient numerical methods
to understand and simulate such problems. It uses the powerful and mathematically rigorous integral
 equation formulation methods for solving multiple scattering problems. 
 Being able to use integral operators allows us to formulate
the solution to a given scattering problem by using traces theorems and variational approaches (see section \ref{sectionMathFormulation}).
 When the boundary $\Gamma$ is general, then boundary element discretization techniques are required
  Ê\cite{AntoineDarbasBook,AntGeuRam10,ColKre83,Mar06,Ned01}. Even if these methods
 are extremely useful for general shapes, they also have some disadvantages.
First, they lead to solving large full linear systems, most particularly when investigating small wavelength problems ($\lambda \ll 1$) and large scatterers
 ($ \textrm{size}(\Omega^{-}) \gg 1$) or collections of many scatterers ($M\gg 1$). These systems require a lot of memory storage and 
 their solution is highly time consuming. The solution can be accelerated by using Krylov subspace solvers
  \cite{AntoineDarbasQJMAM,AntoineDarbasM2AN,AntoineDarbasBook,Saad} in conjunction with fast 
 matrix-vector products
 algorithms (for example Multilevel Fast Multipole Methods \cite{Greengard-Rokhlin} or other compression techniques \cite{AntGeuRam10,GeuzaineBrunoReitich}) 
 but at the price of a loss of accuracy/stability. Second,  
even if boundary element methods provide an accurate solution,   the precision is limited since linear finite element spaces are used
as well as low-order surface descriptions. Going
to higher order basis functions is very complicated and time consuming, most particularly when one wants to integrate with high accuracy
(hyper)singular potentials that are involved in an integral formulation. 

When the geometry is more trivial, then further  simplifications can be realized in the integral equation methods. Indeed, for example,
analytical expressions of the integral operators can be obtained, and spectrally accurate and fast solutions can be derived.
This is the case when considering a disk Ê\cite{AntGeuRam10,Mar06}.  
The Matlab toolbox $\mu$-diff considers the case of a collection of $M$ homogeneous circular cylinders where
Fourier basis expansions can be used (see section \ref{MuDiffFormulation}). Even if disks can be considered as simple geometries, 
a reliable and highly accurate solution is required for wave propagation 
problems (acoustics, electromagnetics, optics, nanophotonics, elasticity) that  involve many circular scatterers, modeling structured or disordered media, 
most particularly when $k$ and $M$ are large (see e.g. \cite{ISI:000253549800023,CassierHazard,ISI:000259270600074,ISI:000249155100055,ISI:000263911800054,MEhrhardtBook, ISI:000256469800019,ISI:000253764200069,HuLu,Joan,KharlamovFilip,ISI:000249786400042,Natarov,PashaevYilmaz,PhysRevA.90.023839,Tsang,TuluYilmaz,NaturePhotonics}). Let us note that all the developments 
in this paper directly apply to 2D TM/TE electromagnetic scattering waves \cite{Mar06} Êeven if our presentation is more related to acoustics.
Furthermore, since multiple scattering is a highly complex  problem with unusual properties, it is desirable to have
a simple modeling tool that helps  to understand the physics properties of such structures. Finally, having a reference solution method for multiple scattering 
leads to the possibility of evaluating the accuracy and performance of other more general numerical methods
like finite element or general integral equation solvers. 
The  goal of the $\mu$-diff Matlab toolbox is to contribute to these different questions.

The structure of the paper is the following. In section \ref{sectionMathFormulation}, we describe the basics of integral operators that are used in $\mu$-diff
and review the most standard integral equation formulations when one wants
to solve the sound-soft scattering problem. In Section \ref{MuDiffFormulation}, we explain the approximation method that
is used in $\mu$-diff to solve the integral equation problems through Fourier series expansions and how to formulate post-processing
data (near- and far-fields for example). In section \ref{sectionFiniteDimensional}, we describe the finite-dimensional approximation
leading to concrete linear systems. Some numerical aspects of the resolution methods are also discussed.
Section \ref{sectionmudiff} details the structure of the $\mu$-diff code and the main functions that are included. 
To illustrate the use of $\mu$-diff, we provide in sections \ref{ExampleSimple} and \ref{sec:penetrable}
 some numerical examples for direct multiple scattering problems 
 (sound-soft, sound-hard, penetrable scatterers).
In addition, we consider  in section \ref{sectionDORT} a more advanced example related to the DORT method (Time Reversal method) in
the presence of homogeneous penetrable
 circular scatterers. All the related files are available in the $\mu$-diff package when downloaded and the simulations can be  reproduced.
Finally, we conclude in section \ref{sectionConclusion}.

\section{Basic theory behind $\mu$-diff: integral equations and formulations for 2D scattering problems}\label{sectionMathFormulation}

\subsection{Definitions and basics on integral operators for scattering}\label{sectionEI}

Let $G$ be the two-dimensional free-space Green's function defined by
$$
\forall \xx, \yy \in \Rb^2, \xx \neq \yy, \quad G(\xx,\yy) = 
\dsp{\frac{i}{4}H_0^{(1)}(k\|\xx-\yy\|)}.\\[0.2cm]
$$
The function $H_0^{(1)}$ is the first-kind Hankel function of order zero.
Integral equations are essentially based upon the Helmholtz integral representation formula \cite[Theorems 3.1 and 3.3]{ColKre83}.
\begin{prop}\label{theo:RepInt}
If $v$ is a solution to the Helmholtz equation in an unbounded connected domain  $\Omega^+$
and satisfies the Sommerfeld radiation condition, then we have
\begin{equation}\label{eqEqInt:RepIntExt}
\int_{\Gamma} - G(\xx,\yy) \dn v(\yy) + \dny G(\xx,\yy) v(\yy) \, \dd\Gamma(\yy) =
\begin{cases}
v(\xx) & \text{ if } \xx \in \Omegaps, \\
0 & \text{otherwise.}
\end{cases}
\end{equation}
If $\vm$ is solution to the  Helmholtz equation in a bounded domain $\Omega^-$, then one gets
\begin{equation}\label{eqEqInt:RepIntInt}
\int_{\Gamma} -G(\xx,\yy) \dn \vm(\yy) + \dny G(\xx,\yy) \vm(\yy) \, \dd\Gamma(\yy) =
\begin{cases}
0 & \text{ if } \xx \in \Omegaps, \\
-\vm(\xx) & \text{ otherwise.}
\end{cases}
\end{equation}
\end{prop}
The integrals on  $\Gamma$ must be understood as duality brackets between the Sobolev space $H^{1/2}(\Gamma)$ and its
 dual space $H^{-1/2}(\Gamma)$. Nevertheless, when the incident wavefield $\uinc$ and the curve $\Gamma$ are sufficiently smooth,
 the scattered field is then regular and the duality bracket can be identified (this  is systematically the case  in the presentation)
  to the (non hermitian) inner product  in $L^{2}(\Gamma)$
$$
 \PSdemi{f}{g} = \int_\Gamma f g d\Gamma.
$$

Let us now introduce the volume single- and double-layer integral operators, respectively denoted by $\Lop$ and $\Mop$, and defined by:
$\forall \xx \in \Rb^2\backslash \Gamma$
$$
\begin{array}{ccc}
\Lop :  \rho &  \longmapsto  &\dsp{\Lop\rho(\xx) = \int_\Gamma G(\xx,\yy) \rho(\yy) \, \dd\Gamma(\yy)},\\
\Mop :  \lambda & \longmapsto & \dsp{\Mop\lambda(\xx) = -\int_\Gamma \dny G(\xx,\yy) \lambda(\yy) \, \dd\Gamma(\yy)}.
\end{array}
$$
We can then express the wavefields $v$  and $\vm$ (see equations \ref{eqEqInt:RepIntExt} and  \ref{eqEqInt:RepIntInt}) as
$$
\begin{cases}
v(\xx) = -\Lop(\dn v|_\Gamma)(\xx) - \Mop (v|_\Gamma)(\xx), & \forall \xx \in \Omegaps, \\[0.3cm]
\vm(\xx) = \Lop(\dn \vm|_\Gamma)(\xx) + \Mop (\vm|_\Gamma)(\xx), & \forall \xx \in \Omegam.
\end{cases}
$$
Furthermore, the single- and double-layer integral operators provide some outgoing solutions to the 
 Helmholtz equation \cite{ColKre98}.
\begin{prop}\label{propEqInt:potentiel}
For any densities $\rho \in \hmdemi$ and $\lambda \in \hdemi$, the functions $\Lop \rho$ and  $\Mop \lambda$ are some outgoing 
solutions to the Helmholtz equation in $\Rb^2 \backslash \Gamma$.
\end{prop}

We now recall the expressions of the trace and normal derivative trace of the volume single- and double-layer potentials which are commonly called
jump relations \cite[Theorem 3.1]{ColKre98}.
\begin{prop}\label{propEqInt:trace}
For any $\xx$ in $\Gamma$, the trace and normal derivative traces of the operators $\Lop$ and $\Mop$ are given by the following relations
 (the signs indicate that $z$ tends towards $x$ from the exterior or the interior of  $\Gamma$) 
\begin{equation}\label{eqEqInt:trace}
\begin{array}{l}
\dsp{\lim_{\zz\in \Omega^{\pm} \to \xx} \Lop \rho (\zz) =L \rho (\xx)}, \hspace{3.0cm}
\dsp{\lim_{\zz\in \Omega^{\pm} \to \xx} \Mop\lambda(\zz) = \left(\mp \frac{1}{2}I + M\right) \lambda(\xx)},\\
\dsp{\lim_{\zz\in \Omega^{\pm} \to \xx} \dnz \Lop \rho (\zz) = \left( \mp \frac{1}{2}I + N\right)\rho(\xx)}, \quad
\dsp{\lim_{\zz\in \Omega^{\pm} \to \xx} \dnz \Mop\lambda(\zz) = D \lambda(\xx)},
\end{array}
\end{equation}
where $I$ is the identity operator, for $\xx \in \Gamma$,
$$
\begin{array}{l l}
\dsp{L\rho(\xx)  =  \int_{\Gamma} G(\xx,\yy) \rho(\yy) \dd\Gamma(\yy)}, \hspace{3.2cm}
\dsp{M\lambda(\xx)  =  -\int_{\Gamma} \dny G(\xx,\yy)\lambda(\yy) \dd\Gamma(\yy)},\\
\dsp{N\rho(\xx) =  \int_{\Gamma} \dnx G(\xx,\yy) \rho(\yy) \dd\Gamma(\yy) = -M^* \rho (\xx)}, \quad
\dsp{D\lambda (\xx)  =  -\dnx \int_{\Gamma} \dny G(\xx,\yy)\lambda (\yy) \dd\Gamma(\yy)}.
\end{array}
$$
\end{prop}
Throughout the paper, the boundary integral operators are  denoted by a roman letter (e.g. $L$)
while the volume integral operators use a calligraphic letter (e.g. $\Lop$). The operator
 $M^* = -N$ is the adjoint operator of $M$, that is
$$
\PSdemi{g}{Mf} = \PSdemi{-Ng}{f}, \qquad \qquad \forall (f,g)\in H^{1/2}(\Gamma)\times H^{-1/2}(\Gamma).
$$
Other properties like compactness or invertibility of integral operators can also be stated \cite{AntGeuRam10, ColKre83, Ned01}.

\subsection{A few boundary integral equations for the Dirichlet problem}\label{AutresEI}

The aim of this section is to provide without  details the most standard integral equation formulations
for solving the 2D scattering problem with Dirichlet boundary condition. These equations serve as model examples
for explaining the way $\mu$-diff works in sections \ref{sectionmudiff} and \ref{sectionnumericalexamples}. We  refer to \cite{AntoineDarbasBook, ThierryThesis, Thi14} 
for further explanations concerning the derivation and properties of these integral equations (like
for the well-posedness  and the possible existence of resonant modes).

The first three integral equations presented here are based on a single-layer representation only
$$
u = \Lop \rho.
$$ 
From this representation and by using  the jump relations, it can be proved that the density $\rho$ is equal to $(-\dn u - \dn \uinc)|_\Gamma$
 and thus has a physical meaning. 
The first integral equation, which is usually called Electric Field Integral Equation (EFIE),  is based on the trace of the single-layer operator
\begin{equation}\label{EFIED}
 L \rho =- \uincg.
\end{equation}
The equation is well-posed and equivalent to the exterior scattering problem (\ref{eqEqInt:ProblemeU}) as soon as $k$ is not an irregular
interior frequency of the associated Dirichlet boundary-value problem \cite{AntoineDarbasBook,ThierryThesis}.

 A second equation, designated by Magnetic Field Integral Equation (MFIE), is
$$
\MFIED \rho =- \duincg.
$$
It is also well-posed and equivalent to the exterior scattering problem (\ref{eqEqInt:ProblemeU}) if
 $k$ is not an interior Neumann resonance \cite{AntoineDarbasBook,ThierryThesis}.  
 
 To avoid the interior resonance problem, Burton and
  Miller \cite{AntoineDarbasBook,BurMil70,ThierryThesis}  consider a linear combination of the EFIE and MFIE.
Let $\alpha$ be a real-valued parameter such that: $0 <\alpha <1$, and 
 $\eta$ be a complex number which satisfies $\Im(\eta) \neq 0$,
where $\Im(\eta)$ is the imaginary part of $\eta$ (the real part is $\Re(\eta)$). Then,   the Combined Field Integral Equation (CFIE)
 \cite{AntoineDarbasBook,HarringtonMautz,ThierryThesis} (also called Burton-Miller integral equation) is given by
$$
\CFIED \rho = - \left[ (1-\alpha) \dn\uinc|_\Gamma + \alpha \eta \uinc|_\Gamma \right].
$$
This integral equation is well-posed for any wavenumber $k$.

Let us now consider $\eta$ as a complex-valued parameter with non zero imaginary part. Then, a fourth
integral representation is based on a linear combination of the single- and double-layer potentials
$$
\ut = -(\eta \Lop + \Mop) \psi + \uinc,
$$
where the total wavefield is defined by $\ut:= u + u^{\inc}$.
The resulting integral equation is obtained by taking the trace of the above relation (see equations (\ref{eqEqInt:trace}))
\begin{equation}\label{BWD}
\BWD \psi = - \uincg.
\end{equation}
This equation, called Brakhage-Werner Integral Equation (BWIE)  \cite{BraWer65}, is well-posed for any 
$k$ and is  equivalent to the exterior scattering problem. Finally, let us note that the surface density
 $\psi$ is unphysical unlike for the three previous equations.

When $\Omegam = \bigcup_{p=1}^{\Nscat}\Omegap$ is multiply connected, all the integral operators can be  written by blocks. For example, the single-layer potential $\Lop\rho$ can be expressed as the sum of elementary potentials $$\Lop\rho = \sum_{p=1}^{\Nscat}\Lop_p\rho_p,$$ where $\rho_p = \rho|_{\Gamma_p}$ and
$$
\Lop_p\rho_p(\xx)  =\int_{\Gamma_p}G(\xx,\yy)\rho_p(\yy)\;\dd\xx, \qquad\forall\xx\in\Rb^2\backslash\overline{\Omega_p}.
$$
Another way of writing the  EFIE (\ref{EFIED})  is then
$$
\left(\begin{array}{c c c c}
L_{1,1} & L_{1,2} & \ldots & L_{1,\Nscat}\\
L_{2,1} & L_{2,2} & \ldots & L_{2,\Nscat}\\
\vdots & \vdots & \ddots & \vdots\\
L_{\Nscat,1} & L_{\Nscat,2} & \ldots & L_{\Nscat,\Nscat}\\
\end{array}\right)
\left(\begin{array}{c}
\rho_1\\
\rho_2\\
\vdots\\
\rho_{\Nscat}
\end{array}\right)
=
-\left(\begin{array}{c}
\uinc|_{\Gamma_1}\\
\uinc|_{\Gamma_2}\\
\vdots\\
\uinc|_{\Gamma_{\Nscat}}
\end{array}\right),
$$
where $L_{p,q}\rho_q = (L_q\rho_q)|_{\Gamma_p}$, with $$\forall \xx\in\Gamma, L_q\rho_q(\xx) = \int_{\Gamma_q}G(\xx,\yy)\rho_q(\xx)\;\dd\yy.$$

\section{Spectral formulation used in $\mu$-diff}\label{MuDiffFormulation}
\label{secEqInt:disque}

We consider now  circular cylinders as scatterers. In this situation, we can explicitly compute  the boundary integral
equations in a Fourier basis,  leading therefore to an efficient computational spectral method when used in conjunction
with numerical linear algebra methods (direct or iterative solvers).
        \subsection{Notations and Fourier basis}\label{secEqInt:BaseFourier}

\begin{figure}[h!]
\begin{center}
\def\svgwidth{10cm}
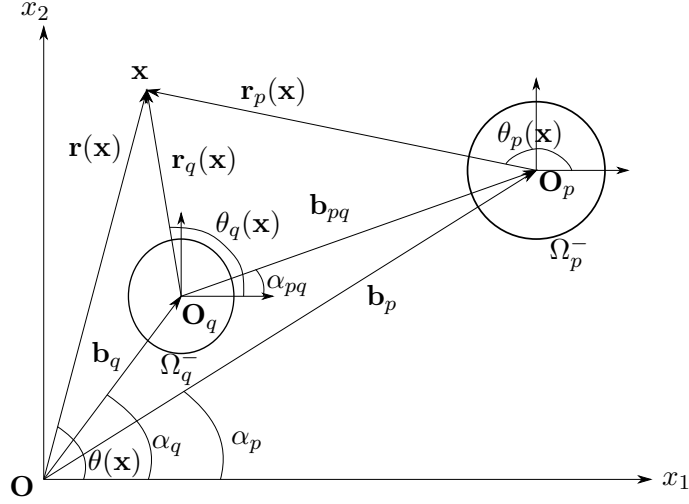
\end{center}
\caption{Illustration of the notations for two disks $\Omegamp$ and $\Omegamq$ and a point $\xxÊ\in \Omegaps$.}
\label{figEqInt:schemanotations}
\end{figure}

Let us consider an orthonormal system $(\OO,\V{\OO x_{1}},\V{\OO x_{2}})$. We assume that the scattering obstacle
 $\Omegam$ is the union of  $M$ disks $\Omegamp$, for $p = 1,\ldots,M$, of radius $a_p$ and center $\OOp$.
 We define $\Gamma_p$ as the boundary of  $\Omegamp$ and by $\dsp{\Gamma = \cup_{p=1 \ldots M}\Gamma_p}$ the boundary of
  $\Omegam$.  The unit normal vector $\nn$ to $\Omegam$ is outgoing. An illustration of the notations is reported on
  Figure \ref{figEqInt:schemanotations}.
  
For any $p=1, \ldots, M$, we introduce $\bbp $ as the vector between the center $\OOp$ and the origin $\OO$
$$
\bbp = \OO \OOp, \qquad \qquad b_p= \left\|\bbp\right\|, \qquad\qquad \alpha_p=Angle(\V{\OO x_{1}}, \bbp),
$$
and, for $q = 1,\ldots,M$, with $q \neq p$, $\bbpq$ as the vector between the centers $\OOq$ and $\OOp$ 
$$
\bbpq = \OOq\OOp, \qquad \qquad b_{pq}=\left\|\bbpq  \right\|, \qquad\qquad
\alpha_{pq}=Angle(\V{\OO x_{1}},\bbpq ).
$$
Furthermore, any point $\xx$ is described by its global polar coordinates
$$
\rr (\xx)= \OO \xx,  \qquad \qquad r (\xx)=\left\|\rr(\xx) \right\|,  \qquad\qquad
\theta(\xx)=Angle(\V{\OO x_{1}},\br(\xx)),
$$
or by its polar coordinates in the orthonormal system associated with the obstacle $\Omegamp$, with $p =1, \ldots, M$, 
$$
\rrp (\xx)= \OOp \xx, \qquad \qquad r_p(\xx)  = \left\|\rrp(\xx)\right\|,  \qquad
\qquad \theta_p(\xx) = Angle(\V{\OOp x_{1}}, \rrp(\xx)).
$$

Let us now build a basis of  $L^2(\Gamma)$ to approximate the integral operators. To this end, we first construct
 a basis of $L^2(\Gamma_p)$ associated with $\Omegamp$, for
 $p=1,\ldots,M$. If the circle $\Gamma_p$ has a radius one and is centered
 at the origin, then a suitable basis of $L^{2}(\Gammap)$ is the spectral Fourier basis of functions $(e^{im\theta})_{m \in \Zb}$.
 We adapt this basis to the general case where $\ap \neq 1$ by introducing, on one hand, the functions $(\phim)_{m\in\Zb}$ defined on $\Rb^{2}$ by: 
$\forall m\in\Zb$, $\forall \xx \in \Rbb$, $\varphi_m(\xx) = e^{im\theta(\xx)}$,
and, on the other hand,  the functions  $(\phimp)_{1\leqslant p \leqslant M, \; m\in\Zb}$  given by
$$
\forall p=1,\ldots,M, \forall m\in\Zb, 
\forall \xx \in \Gamma_p, \qquad\varphi_m^p(\xx) = \frac{\varphi_m(\rr_p(\xx))}{\sqrt{2 \pi a_p}} = \frac{e^{im\theta_p(\xx)}}{\sqrt{2 \pi a_p}}.
$$
For $p=1,\ldots,M$, the family $\dsp{\left( \varphi_m^p \right)_{m \in \Zb}}$ forms an orthonormal basis of
 $L^2(\Gamma_p)$ for the hermitian inner product $\PSGammap{\cdot}{\cdot}$
$$
\forall f,g\in L^{2}(\Gammap),\qquad \PSGammap{f}{g} = \int_{\Gammap} f(\xx) \overline{g(\xx)} \dd\Gammap(\xx).
$$
To build a basis of $L^2(\Gamma)$, we introduce the functions $\Phi_m^p$ of $L^2(\Gamma)$ as the union of these $M$ families 
$$
\forall p,q = 1,\ldots, M, \forall m\in\Zb, \qquad \Phi_m^p |_{\Gamma_q} = 
\begin{cases}
        0 & \text{if } q \neq p, \\
        \varphi_m^p & \text{if } q = p.
\end{cases}
$$
The family $\BF = \{ \Phi_m^p, \, m \in \Zb, p =1, \ldots, M\}$, also called Fourier or spectral basis, is a Hilbert basis of
 $L^2(\Gamma)$ for the usual scalar product $(\cdot,\cdot)_{L^{2}(\Gamma)}$.

\subsection{Integral operators - integral equations for a cluster of circular cylinders}\label{integralinfinite}

In view of a numerical procedure, $\mu$-diff uses the weak formulation of the EFIE (\ref{EFIED}) in 
 $L^{2}(\Gamma)$ based on the Fourier basis $\BF$ 
$$
\begin{cases}
\text{Find $\rho \in H^{-1/2}(\Gamma)$ such that for any $p=1,\ldots,M,$ and $m\in\Zb$,} & \\
\PSGamma{L\rho}{\Phimp} = -\PSGamma{\uincg}{\Phimp}. &
\end{cases}
$$
Since $\uinc$ is assumed to be smooth enough (typically $\Cscr^{\infty}$) and that $\Gamma$ is $\Cscr^{\infty}$, then
the scattered wavefield is also $\Cscr^{\infty}(\Omegaps)$ and the density  $\rho$ is (at least) in
 $H^{1/2}(\Gamma)$. Therefore, $\rho$ can be expanded in  $\BF$ as 
$$
\rho = \sum_{q=1}^{M}\sum_{n\in\Zb}\rhonq\Phinq$$ and the weak form of the EFIE  is 
$$
\begin{cases}
\text{Find the Fourier coefficients $\rhonq\in\Cb$, for $q=1,\ldots,M$, and $n\in\Zb$, such that,} & \\
\dsp{\forall p=1,\ldots,M, \;\forall m\in\Zb, \qquad \sum_{q=1}^{M}\sum_{n\in\Zb}\rhonq\PSGamma{L\Phinq}{\Phimp} = -\PSGamma{\uincg}{\Phimp}}. &
\end{cases}
$$
This formulation can be written under the following matrix form $
\Lbt\Rhot = \UUt$,
where the infinite matrix representation
 $\Lbt =(\Lbtpq)_{1\leqslant p,q\leqslant M}$ and the infinite vectors
  $\Rhot=(\Rhot^{p})_{1\leqslant p\leqslant M}$ and
   $\UUt=(\UUt^{p})_{1\leqslant p\leqslant M}$ are defined by blocks   as
\begin{equation}\label{eqEqInt:MatLbt}
\Lbt =
\left[
\begin{array}{c c c c}
\Lbt^{1,1} & \Lbt^{1,2} & \ldots & \Lbt^{1,M} \\
\Lbt^{2,1} & \Lbt^{2,2} & \ldots & \Lbt^{2,M} \\
\vdots & \vdots & \ddots & \vdots\\
\Lbt^{M,1} & \Lbt^{M,2} & \ldots & \Lbt^{M,M}
\end{array}
\right],
\qquad \qquad
\Rhot =
\left[
\begin{array}{c}
\Rhot^{1} \\
\Rhot^{2} \\
\vdots \\
\Rhot^{M}
\end{array}
\right],\qquad 
\UUt =
\left[
\begin{array}{c}
\UUt^{1} \\
\UUt^{2} \\
\vdots \\
\UUt^{M}
\end{array}
\right],
\end{equation}
with, for any $p,q=1,\ldots,M$, and $m,n\in\Zb$:
$
\Lbtpqmn = \PSGamma{L\Phinq}{\Phimp}$, $\Rhot_{m}^{p} = \rhomp$ and
$\UUt_{m}^{p} = \PSGamma{-\uincg}{\Phimp}$.

For the other integral formulations (section \ref{AutresEI}) or even for any other boundary condition, 
 the expressions of the three boundary integral operators  $M$, $N$ and $D$ are needed.
Therefore, to compute an integral equation, we introduce the infinite matrices $\Mbt = (\Mbtpq)_{1\leqslant p,q\leqslant M}$, $\Nbt= (\Nbtpq)_{1\leqslant p,q\leqslant M}$ and
 $\Dbt= (\Dbtpq)_{1\leqslant p,q\leqslant M}$, with the same block structure as $\Lbt$ (see equation
  (\ref{eqEqInt:MatLbt})). For $p,q =1,\ldots, M$, the coefficients of the infinite matrices $\Mbtpq$, $\Nbtpq$ and $\Dbtpq$ are defined for any indices
   $m$ and $n$ in $\Zb$ by
\[
\Mbtpqmn = \PSGamma{M\Phinq}{\Phimp},
\Nbtpqmn = \PSGamma{N\Phinq}{\Phimp}, \textrm{ and }
\Dbtpqmn = \PSGamma{D\Phinq}{\Phimp}.
\]
For a numerical implementation, we can explicitly compute \cite{JACT,ThierryThesis} the matrix blocks
 $\Lbtpq$, $\Mbtpq$, $\Nbtpq$ and $\Dbtpq$ involved in $\Lbt$, $\Mbt$, $\Nbt$ and $\Dbt$, for
  $p,q=1,\ldots,M$. To this end, we introduce the infinite diagonal matrices
   $\Jbtp$, $\dJbtp$, $\Hbtp$ and $\dHbtp$, with general terms, for $m \in \Zb$,
$$
\Jbtp_{mm} = \Jm(ka_{p}), \hspace{0.5cm}
\dJbtp_{mm} = \Jmp(ka_{p}), \hspace{0.5cm}
\Hbtp_{mm} = \Hm(ka_{p}), \hspace{0.5cm}
\dHbtp_{mm} = \Hmp(ka_{p}).
$$
In addition, let $\Ibtp$ be the infinite identity matrix, and, for $q\neq p$, the infinite separation matrix $\Sbtpq$ between the obstacles $\Omegamp$ and
 $\Omegamq$, defined by
$$
\Sbtpq=(\Sbtpqmn)_{m\in \Zb,n\in
\Zb}\qquad \text{and}\qquad \Sbtpqmn=\Smnbpq = H_{m-n}^{(1)} (k b_{pq})e^{i(m-n)\alpha_{bq}}.
$$
Under these notations, we rewrite the blocks $\Lbtpq$, $\Mbtpq$, $\Nbtpq$ and $\Dbtpq$ of the infinite matrices
 $\Lbt$, $\Mbt$, $\Nbt$ and $\Dbt$ under the matrix form, for any $p,q= 1,\ldots,M$,

\qquad\qquad \textbullet $\quad\dsp{\Lbtpq = 
\begin{cases}
\dsp{\frac{i\pi a_p}{2} \Jbtp\Hbtp,} & \text{ if } p = q,\\[0.3cm]
\dsp{\frac{i \pi\sqrt{a_p a_q}}{2} \Jbtp (\Sbtpq)^T\Jbtq,} &\text{ if } p \neq q,
\end{cases}}$\\[0.5cm]

\qquad\qquad \textbullet $\quad\dsp{\Mbtpq = 
\begin{cases}
\dsp{ - \frac{1}{2}\Ibtp - \frac{i \pi k a_p}{2} \Jbtp \dHbtp = \frac{1}{2}\Ibtp - \frac{i \pi k a_p}{2} \dJbtp \Hbtp,} &\text{ if } p = q,\\[0.3cm]
\dsp{- \frac{i k \pi\sqrt{a_p a_q}}{2} \Jbtp (\Sbtpq)^T \dJbtq,}&\text{ if } p \neq q,
\end{cases}}$\\[0.5cm]

\qquad\qquad \textbullet $\quad\dsp{\Nbtpq =
\begin{cases}
\dsp{\frac{1}{2}\Ibtp + \frac{i \pi k a_p}{2} \Jbtp \dHbtp  = -\frac{1}{2}\Ibtp + \frac{i \pi k a_p}{2} \dJbtp \Hbtp,} &\text{ if } p = q, \\[0.3cm]
\dsp{\frac{i k \pi\sqrt{a_p a_q}}{2} \dJbtp (\Sbtpq)^T \Jbtq,}& \text{ if } p \neq q, 
\end{cases}}$\\[0.5cm]

\qquad\qquad \textbullet $\quad\dsp{\Dbtpq =
\begin{cases}
\dsp{\frac{i \pi k^2 a_p}{2} \dJbtp\dHbtp,} & \text{ if } p = q, \\[0.3cm]
\dsp{ - \frac{i k^2 \pi\sqrt{a_p a_q}}{2} \dJbtp (\Sbtpq)^T \dJbtq,} &\text{ if } p \neq q,
\end{cases}}$\\[0.5cm]
where $(\Sbtpq)^T$ is the transpose matrix of the separation matrix $\Sbtpq$. 

The integral equations involve the trace or normal derivative trace of the incident wavefield on $\Gamma$.
We  have already introduced the infinite vector $\UUt$ of the coefficients of $\uincg$ in the Fourier basis.
We then define similarly the infinite vector $\dUUt = (\dUUt^{p})_{1\leqslant p \leqslant M}$ of the coefficients of the normal derivative trace
 $\duincg$, such that
$$
\forall p=1,\ldots,M,\;\forall m\in\Zb,\qquad \dUUtmp = \PSGamma{\duincg}{\Phimp}.
$$
Finally, the density changes according to the integral equation and most particularly with respect to the boundary condition. 
To keep the same notations as previously, we introduce the densities
 $\lambda$ and $\psi$ (used in the BWIE) that are expanded in the Fourier basis as
$$
\lambda = \sum_{p=1}^{M}\sum_{m\in\Zb} \lambdamp\Phimp \qquad \text{ and }\qquad \psi = \sum_{p=1}^{M}\sum_{m\in\Zb} \psimp\Phimp.
$$
Finally, we set: $\lambdabt = (\lambdabt^{p})_{1\leqslant p \leqslant M}$ and
 $\Psit = (\Psit^{p})_{1\leqslant p \leqslant M}$, where each block  $\lambdabt^{p} = (\lambdabt_{m}^{p})_{m\in\Zb}$ and
  $\Psit^{p} = (\Psit_{m}^{p})_{m\in\Zb}$ is defined by: $\forall m\in\Zb$,
$\lambdabt_{m}^{p} = \lambdamp$ and $\Psit_{m}^{p} = \psimp$.

        \subsection{Projection of the incident waves in the Fourier basis}\label{secEqInt:SecondMembre}

To fully solve one of the integral equations (EFIE, MFIE, CFIE or Brakhage-Werner), we need to compute the Fourier coefficients of the
trace and normal derivative traces of the incident wave. We give the results  for both an incident plane wave 
and a pointwise source term (Green's function).

For an incident plane wave, the following proposition holds \cite{AntChnRam08}.
\begin{prop}\label{propEqInt:UincFourier}
Let us assume that $\uinc$ is an incident plane wave of direction $\Beta$, with $\Beta = (\cos(\beta),\sin(\beta))$ and
 $\beta\in[0,2\pi]$, i.e.
$$
\forall\xx\in\Rb^{2},\qquad\uinc(\xx) = e^{ik\Beta\cdot\xx}.
$$
Then we have the following equalities
$$
\UUt_{m}^{p}= \PSGamma{\uincg}{\Phimp} = \dmp \Jm(ka_p), \qquad  \dUUtmp=\PSGamma{\dn\uincg}{\Phimp} = k \dmp \Jmp(ka_p),
$$
with
$
\dmp = \sqrt{2\pi a_p} e^{ik \Beta \cdot \bbp} e^{i m (\pi/2 - \beta)}
$.
\end{prop}

Let us consider now an incident wave emitted by a pointwise source located at $\ssb\in\Omegaps$, i.e.
the wave $\uinc$ is the Green's function centered at $\ssb$.
The  Fourier coefficients  of the trace and normal derivative trace of $\uinc$ on $\Gamma$  are then given
by the following proposition \cite{ThierryThesis}.
\begin{prop}\label{propEqInt:UincGreenFourier}
Let $\ssb\in\Omegaps$. We assume that the incident wave $\uinc$ is the Green's function centered at $\ssb$ 
$$
\forall \xx\in\Rb^{2}\setminus\{\ssb\},\qquad \uinc(\xx) = G(\xx,\ssb) = \frac{i}{4}\Hz(k\|\xx-\ssb\|).
$$
The Fourier coefficients in $\Bscr$ of the trace and normal derivative trace of the incident wave on $\Gamma$ are respectively given by
$$
\UUt_{m}^{p} = \PSGamma{\uincg}{\Phimp} = \frac{i \pi\ap}{2}\Jm(ka_p)\Hm(k\rp(\ssb))\overline{\Phimpt(\ssb)}
$$
and
$$
\dUUtmp = \PSGamma{\dn\uincg}{\Phimp} = k \frac{i \pi\ap}{2} \Jmp(k \ap) H_{m}^{(1)}(k \rp(\ssb)) \overline{\Phimpt(\ssb)}.
$$
\end{prop}

	\subsection{Near- and far-fields evaluations}\label{secEqInt:quantite}


By using the Graf's addition theorem \cite{Mar06,ThierryThesis}, we can compute the expression of the single- and double-layer potentials
at a point  $\xx$ located in the propagation domain $\Omegaps$.
\begin{prop}\label{propEqInt:LMphi}
Let $\rho\in L^{2}(\Gamma)$ and $\mu\in H^{1/2}(\Gamma)$ be two densities admitting the following decompositions in the Fourier basis $\BF$ 
$$
\rho = \sum_{p=1}^{M}\sum_{m\in\Zb} \rhomp\Phimp \qquad\text{ and }\qquad \lambda = \sum_{p=1}^{M}\sum_{m\in\Zb} \lambdamp\Phimp.
$$
Then, for any point $\xx$ in the  domain of propagation $\Omegaps$, the single-layer potential reads
$$
\Lop\rho(\xx) = \sum_{p=1}^M\sum_{m\in\Zb}\rhomp \Lop \Phimp(\xx) =\sum_{p=1}^M\sum_{m\in\Zb}\rhomp \frac{i\pi a_p}{2} J_m(ka_p)\Hm(kr_p(\xx)) \Phimpt(\xx),
$$
and the double-layer potential can be expressed as
$$
\Mop\lambda(\xx) = \sum_{p=1}^M\sum_{m\in\Zb}\lambdamp\Mop \Phimp(\xx) = -\sum_{p=1}^M\sum_{m\in\Zb} \lambdamp\frac{i\pi ka_p}{2} J_m'(ka_p)\Hm(kr_p(\xx)) \Phimpt(\xx).
$$
\end{prop}
Proposition \ref{propEqInt:LMphi} implies that, for any  $\xx$ in $\Omegaps$,
$$
 u(\xx) = \Lop\rho(\xx) + \Mop\lambda(\xx) = \sum_{p=1}^M\sum_{m\in\Zb}\frac{i\pi a_p}{2} \left[\rhomp  J_m(ka_p) + \lambdamp \Jmp(k\ap)\right]\Hm(kr_p(\xx)) \Phimpt(\xx).
$$


For computing the far-field pattern, let us recall that the scattered field $u$ admits the following Helmholtz's integral representation:
$u = \Lop \rho + \Mop \lambda$,
where $\rho$ and $\lambda$ are two unknown densities. In the polar coordinates system $(r,\theta)$ and by using an asymptotic expansion
of  $u$ when $r\to +\infty$, the following relation holds \cite{ColKre83}
$$
\forall \theta\in [0,2\pi],\qquad u(r,\theta) = \frac{e^{ikr}}{r^{1/2}} \left[ a_{\Lop}(\theta) + a_{\Mop}(\theta) \right] + \GrandO{\frac{1}{r^{3/2}}},
$$
where $a_{\Lop}$ and $a_{\Mop}$ are the radiated far-fields  for the single- and double-layer potentials, respectively, defined for any angle
 $\theta$ of $[0,2\pi]$ by
$$
\begin{cases}
\dsp{a_{\Lop}(\theta) = \frac{1}{\sqrt{8k\pi}}e^{i\pi /4} \int_{\Gamma} e^{-ik \thetab \cdot \yy} \rho(\yy) \dd \Gamma(\yy),}\\[0.4cm]
\dsp{a_{\Mop}(\theta) = \frac{1}{\sqrt{8k\pi}}e^{i\pi /4} \int_{\Gamma} -\frac{ik}{\|\yy\|} \thetab \cdot \yy e^{-ik \thetab \cdot \yy} \lambda(\yy) \dd \Gamma(\yy),}
\end{cases}
$$
with $\thetab:=(\cos(\theta),\sin(\theta))$. In addition, the Radar Cross Section (RCS) is defined by
$$
\forall \theta\in[0,2\pi], \quad \textrm{RCS}(\theta) = 10\log_{10}\left(2\pi\Abs{\aLop(\theta) + \aMop(\theta)}^{2}\right) (\dB).
$$
To optimize the far-fields computation, these relations can be written thanks to the inner product between two infinite vectors.
Indeed, let us introduce $\aabt_{\Lop} = ((\aabt_{\Lop})^{p})_{1\leqslant p \leqslant M}$ and $\aabt_{\Mop} = ((\aabt_{\Mop})^{p})_{1\leqslant p \leqslant M}$,
where $(\aabt_{\Lop})^{p}$ and $(\aabt_{\Mop})^{p}$ are given by:  $\forall p=1,\ldots,M$,
$$
\left\{\begin{array}{l}
\dsp{(\aabt_{\Lop})^{p} = \Big((\aabt_{\Lop})^{p}_{m}\Big)_{m\in\Zb}, \qquad (\aabt_{\Lop})^{p}_{m} = \frac{ie^{-i\pi/4}\sqrt{a_{p}}}{2\sqrt{k}} e^{-i\bp k\cos(\theta-\alphap)}\Jm(k\ap)e^{im(\theta-\pi/2)},}\\[0.3cm]
\dsp{(\aabt_{\Mop})^{p} = \Big((\aabt_{\Lop})^{p}_{m}\Big)_{m\in\Zb}, \qquad (\aabt_{\Lop})^{p}_{m} = \frac{ie^{-i\pi/4}\sqrt{ka_{p}}}{2} e^{-i\bp k\cos(\theta-\alphap)}\Jmp(k\ap)e^{im(\theta-\pi/2)}.}
\end{array}\right.
$$
Then, we obtain the following: $\dsp{a_{\Lop}(\theta) = (\aabt_{\Lop})^{T}\Rhot}$ and $\dsp{a_{\Mop}(\theta) = (\aabt_{\Mop})^{T}\lambdabt}$.

	\section{Finite-dimensional approximations and numerical solutions proposed in $\mu$-diff}\label{sectionFiniteDimensional}

We now have all the ingredients to numerically solve the four integral equations EFIE, MFIE, CFIE and BWIE,
for sound-soft obstacles. In fact, any integral equation for any boundary condition can be solved according to the previous developments.
In practice, the infinite Fourier systems need to be truncated to get a finite dimensional problem: we must pass from
a sum over  $m\in \mathbb{Z}$ to a finite number of Fourier modes that depends on $ka_{p}$, $p=1,...,M$.
Let us consider e.g. the EFIE, the extension to the other boundary integral operators being direct. The EFIE is given by
 equation (\ref{eqEqInt:MatLbt}): $\Lbt\Rhot = -\UUt$. To truncate each Fourier series associated with 
  $(\Phimp)_{m\in\Zb}$ for the obstacle $\Omegamp$, we only keep $2N_{p} +1$ modes in such a way that the indices $m$ of the truncated series
  satisfy: $\forall p=1,\ldots,M$, $-\Np\leqslant m \leqslant \Np$. The truncation parameter $N_{p}$ must be fixed large enough, with
  $N_{p} \geqslant k a_{p}$, for $p=1,...,M$. An example \cite{AntChnRam08,JACT} is: $N_{p} = k a_{p} +ÊC_{p}$, where
   $C_{p}$ weakly grows with $k a_{p}$. A numerical study of the parameter $\Np$ is proposed in \cite{AntChnRam08,JACT} where 
   the following formula leads to a stable and accurate computation
\begin{equation}\label{eq:Np}
 \Np = \left[ k a_p+ \left(\frac{1}{2\sqrt{2}} \ln (2\sqrt{2} \pi k a_p \varepsilon^{-1})\right)^{\frac{2}{3}} (k a_p)^{1/3} +1\right ],
\end{equation}
where $\varepsilon$ is a small parameter (related to the relative tolerance required in the iterative Krylov subspace solver used
for solving the truncated linear system (\ref{EFIEDimensionFinie}), see  \cite{AntChnRam08,JACT}).

The resulting  linear system writes
\begin{equation}\label{EFIEDimensionFinie}
\Lb\Rho = -\UU,
\end{equation}
where we introduced the block matrix $\Lb = (\Lbpq)_{1\leqslant p,q\leqslant M}$ and the vectors $\Rho = (\Rho^{p})_{1\leqslant p\leqslant M}$ and
 $\UU = (\UU^{p})_{1\leqslant p\leqslant M}$ defined by
\begin{equation}\label{eqEqInt:EFIEtronque}
\Lb = \left[
\begin{array}{c c c c}
\Lb^{1,1} & \Lb^{1,2} & \ldots & \Lb^{1,M} \\
\Lb^{2,1} & \Lb^{2,2} & \ldots & \Lb^{2,M} \\
\vdots & \vdots & \ddots & \vdots\\
\Lb^{M,1} & \Lb^{M,2} & \ldots & \Lb^{M,M}
\end{array}
\right],
\qquad \qquad
\Rho =
\left[
\begin{array}{c}
\Rho^{1} \\
\Rho^{2} \\
\vdots \\
\Rho^{M}
\end{array}
\right],\qquad \qquad
\UU =
\left[
\begin{array}{c}
\UU^{1} \\
\UU^{2} \\
\vdots \\
\UU^{M}
\end{array}
\right].
\end{equation}
For $p,q= 1,\ldots, M$,
 the  complex-valued matrix $\Lbpq$ is of size $(2\Np+1)\times(2N_{q}+1)$ and its coefficients $\Lbpqmn$ are: $\Lbpqmn = \Lbtpqmn$,
 for $m=-\Np,\ldots,\Np$, $ n = -\Nq,\ldots,\Nq$.
The complex-valued components of the vector $\Rho^{p} = (\Rho^{p}_{m})_{-\Np\leqslant m \leqslant\Np}$ of size $2\Np+1$ are
  the approximate Fourier coefficients  $\rhomp$ of $\rho$. For the sake of clarity, we keep on writing: $\Rho^{p}_{m} = \Rhot^{p}_{m} = \rhomp$,
  for all $ m =-\Np,\ldots,\Np$.
The complex-valued vector $\UU^{p} = (\UU^{p}_{m})_{-\Np\leqslant m \leqslant\Np}$ is composed of the  $2\Np+1$ Fourier
coefficients of the trace of the incident wave on $\Gamma$, i.e.  $\UU^{p}_{m} = \UUt^{p}_{m} = \PSGamma{\uincg}{\Phimp}$,
$\forall m =-\Np,\ldots,\Np$.
If $\Ntot = \sum_{p=1}^{M}(2\Np +1)$ denotes the total number of modes, the size of the complex-valued matrix
 $\Lb$ is then $\Ntot\times\Ntot$. More generally, all the boundary integral operators can be truncated according to this process.
 Concerning the notations, it is sufficient to formally omit the tilde symbol $\sim$ over the quantities involved in sections
  (\ref{integralinfinite})-(\ref{secEqInt:quantite}).

Since  the four finite-dimensional matrices $\Lb$, $\Mb$, $\Nb$ and $\Db$ that respectively correspond to the four
boundary integral operators $L$, $M$, $N$ and $D$ can be computed,  the linear systems that approximate the EFIE, MFIE, CFIE and
 BWIE can be stated. For example, the CFIE leads to (with $0\leqslant \alpha\leqslant 1$ and $\Im(\eta)\neq 0$)
\begin{equation}\label{eqEqInt:CFIEDapp}
\left[ \alpha\eta\Lb + (1-\alpha) \left(\frac{\Ib}{2} + \Nb\right)\right]\Rho = -\alpha\eta\UU - (1-\alpha)\dUU.
\end{equation}
Let us remark that the matrix obtained after discretization is always a linear combination of the four integral operators
 $\Lb$, $\Mb$, $\Nb$, $\Db$ and the identity matrix  $\Ib$. As a consequence, for a given integral equation,
 the resulting matrix is of size $\Ntot\times\Ntot$ and has the same block structure as e.g. 
  $\Lb$ (see equation  (\ref{eqEqInt:EFIEtronque})).
 The finite-dimensional linear system (\ref{EFIEDimensionFinie}) (or (\ref{eqEqInt:CFIEDapp})) is accurately solved in $\mu$-diff
by using the Matlab direct solver or a preconditioned Krylov subspace linear solver 
that uses fast matrix-vector products based on Fast Fourier Transforms (FFTs), the choice of the linear algebra strategy
(direct vs. iterative) depending on the configuration with respect to $ka_{p}$ and  
$M$. The preconditioner included in $\mu$-diff is based on the diagonal of the integral operator matrix representation
which is solved and corresponding to single scattering.
The use of FFTs is made possible since the off-diagonal blocks of the integral operators can be written as the products
of diagonal and Toeplitz matrices \cite{AntChnRam08,JACT} (see e.g. the matrices $\Sbtpqmn$ in section \ref{integralinfinite}). In addition,
low memory is only necessary when $ka_{p}$ is large enough since  the storage of the Toeplitz
matrices can be optimized. This resulting storage technique is called \textit{sparse} representation in $\mu$-diff,
 in contrast with the \textit{dense} (full) storage of the complex-valued matrices.
 Let us assume that $a_p \approx a$, for $1 \leqslant p \leqslant M$.
In terms of storage, the dense version of a matrix requires to store about $4M^{2}[ka]^{2}$ coefficients (assuming that $N_{p}$ are fixed
by formula (\ref{eq:Np}), and $[r]$ denotes the integer part of a real number $r$) while the sparse storage needs about
$4M^{2}[ka]$ complex-valued coefficients. In terms of computational time for solving the linear system,
the direct (multithreaded) gaussian solver included in Matlab leads to a cost that scales with $\mathcal{O}(M^{3}(ka)^{3})$. For the preconditioned iterative Krylov subspace
methods (i.e. restarted GMRES)), the global cost is $\mathcal{O}(M^2 ka \log_{2}(ka))$, the converge rate depending on
the physical situation and robustness of the preconditioner. From these remarks, we deduce that an iterative method is an efficient  and cheap 
alternative to a direct solver for large wavenumbers $ka$, but also for large $M$.
 We refer to \cite{AntChnRam08,JACT} for a thorough computational study of the various numerical strategies.
A few examples in $\mu$-diff are provided (see section  \ref{sectionnumericalexamples} and the corresponding scripts) with the toolbox.
Finally, the post-processing formulas (near- and far-fields quantities) clearly inherits of the truncation procedure (see section \ref{secEqInt:quantite}).

\section{Structure  of the $\mu$-diff Matlab toolbox}\label{sectionmudiff}

Because $\mu$-diff includes all the integral operators that are needed in scattering (traces and normal derivative
traces of the single- and double-layer potentials), a large class of scattering problems can be solved. 
Concerning the geometrical configurations, any deterministic or random distribution of disks is possible. Finally,
$\mu$-diff includes post-processing facilities like \textit{e.g.}: surface and far-fields computations, total and scattered exterior (near-field)
visualization...

We now introduce the $\mu$-diff Matlab toolbox by explaining the main predefined functions and their relations
with the previous mathematical derivations.
Section \ref{sectionPreProcessing}  shows how to define the scattering configuration (geometry and physical parameters). Section
 \ref{ResolutionMuDiff}  presents the way the integral equations must be defined and solved.
  Finally,
 section \ref{sectionPostProcessing} describes the data post-processing. 
 To be concrete, we propose to fully treat  in section \ref{ExampleSimple} 
the example of multiple scattering by a collection of randomly distributed sound-soft and sound-hard circular cylinders based on the EFIE.
Section \ref{sec:penetrable} presents an example of scattering by penetrable obstacles and a more advanced example is considered in section
 \ref{sectionDORT} for time reversal in homogeneous media.

The $\mu$-diff toolbox  is organized following the five subdirectories:
\begin{itemize}
\item \texttt{mudiff/PreProcessing/}: pre-processing data functions (incident wave and geometry) (section \ref{sectionPreProcessing}).
\item \texttt{mudiff/IntOperators/}: functions for the four basic integral operators (dense and sparse structure)
used in the definition of the integral equations to solve (section \ref{ResolutionMuDiff}). 
\item \texttt{mudiff/PostProcessing/}: post-processing functions of the solution (trace and normal derivative traces,
computation of the scattered/total wavefield 
at some points of the spatial domain or on a grid, far-field and RCS) 
(section \ref{sectionPostProcessing}).
\item \texttt{mudiff/Common/}: this directory includes functions that are used in $\mu$-diff but which does not need to be known
from the standard user point of view.
\item \texttt{mudiff/Examples/}: various scripts are presented for the user in standard configurations.
\end{itemize}
In addition, the $\mu$-diff user-guide can be found under the directory \texttt{mudiff/Doc/}.

\subsection{Pre-processing: physical and geometrical configurations}\label{sectionPreProcessing}
All the pre-processing functions are included in the directory \texttt{mudiff/PreProcessing/}. 

The pre-processing (\texttt{mudiff/PreProcessing/IncidentWave}) in $\mu$-diff consists first in defining the scattering parameters (incidence angle $\beta$ or location of the point source, wavenumber $k$). This provides the possibility of defining the traces and normal derivative traces 
of the incident wavefield through the global function \texttt{IncidentWave}
(plane wave or point sources), or through the specific functions \texttt{PlaneWave}, \texttt{DnPlaneWave} (plane wave), \texttt{PointSource},  \texttt{DnPointSource}
(point source) in view of writing any integral formulation. The global function also allows to build a  vector mixing the
trace and normal derivative trace of an incident wave (e.g. a vector combining  \texttt{PlaneWave} and \texttt{DnPlaneWave}). Let us also note that the user could define is own incident field in the Fourier 
basis by sampling the signal.

Next, the geometrical configuration can be described thanks to functions available in the directory \texttt{mudiff/PreProcessing/Geometry}.
The user can define himself the centers and radii (\texttt{(O, a)}) of the circular cylinders,   create a rectangular (\texttt{RectangularLattice} function)
 or triangular (\texttt{TriangularLattice} function) lattice of circular cylinders  or can even build a random set of cylinders in a rectangular domain
 (\texttt{CreateRandomDisks} function), specifying many geometrical parameters to describe dilute or dense random media
 (minimal and maximal size of the disks, minimal distance between each disk,\ldots) and even create holes in the domain where no disk must overlap
 (this can be interesting for example for numerically building photonics crystals with cavity).

\subsection{Defining and solving an integral equation}\label{ResolutionMuDiff}
The functions defining the integral operators are in the directory \texttt{mudiff/IntOperators/} which has the \texttt{Dense/}
and \texttt{Sparse/} subdirectories for the dense (matrix) and sparse (@function) representations of the four basic integral operators used in scattering,
i.e. $\Lb$, $\Mb$, $\Nb$ and $\Db$. Preconditioned versions of the operators by their diagonal part are also defined (based on single scattering \cite{AntChnRam08,JACT}).
For example, for a Dirichlet boundary value problem, the EFIE (\ref{EFIED}), which is based on a single-layer representation, can be built by using the function \texttt{SingleLayer} for a dense matrix version or the function \texttt{SpSingleLayer} to get a sparse representation. 
Nevertheless, from the user point of view, there is no need to enter into the detail of all the related functions. Indeed, a frontal function,
called \texttt{IntegralOperator}, allows to directly build a linear combination of the previous integral operators, which are all indexed by a hard-coded number. This provides a very convenient way
when one does not want to use the specific functions or need to build a more complicated operator. For example, the spectral (dense) construction of the BWIE for the Dirichlet problem can be written
\[
\texttt{IntegralOperator(O, a, M\_modes, k, [1, 2, 3], [0.5, -eta, -1]);}
\]
for
\begin{equation}\label{BWDMuDiff}
 \frac{1}{2}\Ib-\eta \Lb - \Mb.
\end{equation}
The vector \texttt{M\_modes} is such that \texttt{M\_modes(p)}$=N_p$, the argument vector \texttt{[1, 2, 3]} refers to respectively the operators Identity 
($\texttt{1}$), $L$ ($\texttt{2}$) and $M$ ($\texttt{3}$)
 and the last one \texttt{[0.5, -eta, -1]} carries the weight to apply to each operator in the linear combination (\texttt{eta} must previously
have taken a prescribed complex value in the script). Without entering too much into details, each block of the final global matrix can be specified thanks to this numbering (instead of a vector, a 2D- or a 3D-array is then 
considered as argument). For the sparse version, the operators are stored using the Matlab \texttt{cell} structure. Building a linear combination of the
different integral operators is then slightly different: each operator is assembled separately and all the integral operators are next combined
 during the sparse matrix-vector product as shown below.

Once the dense or sparse integral operator has been defined and the right-hand side has been computed, then the integral equation can be solved.
For the dense representation of the integral operator, it is possible to use a direct Gauss solver
 (based on the  backslash $\backslash$ Matlab operator)  or any iterative Krylov subspace
  solver available in Matlab (GMRES, BiCGStab,\ldots). When the sparse structure
 is used, there is no other possibility than using an iterative solver. For the BWIE, the following syntax is required to build
 the function representing the integral operator (\ref{BWDMuDiff}) which is next called 
 for solving the equation (\ref{BWD}) by using the restarted GMRES Matlab solver
   \[\begin{array}{l}
 \texttt{Uinc = PlaneWave(O, a, M\_modes, k, beta\_inc);}\\ 
 \texttt{SpI = SpIdentity(O, a, M\_modes);}\\ 
 \texttt{SpL = SpSingleLayer(O, a, M\_modes, k);}\\
 \texttt{SpM = SpDoubleLayer(O, a, M\_modes, k);}\\
 \texttt{[psi\_SpBW,FLAG\_SpBW,RELRES\_SpBW,ITER\_SpBW,RESVEC\_SpBW] =...}\\
\hspace{2cm}Ê \texttt{gmres(@(X)SpMatVec(X,M\_modes, \{SpI, SpL, SpM\}, [0.5, -eta\_BW, -1]),... }\\
\hspace{5cm}Ê \texttt{Uinc, RESTART, TOL, MAXIT, [], []);}
\end{array}
\]
Let us note that the way $\mu$-diff is built allows to define the matrices and vectors block-by-block and thus to solve any integral equation formulation which can for example
take into account different boundary conditions on the circular cylinders, complex wavenumber for the interior/exterior of a disk,\ldots

\subsection{Post-processing of computed outputs}\label{sectionPostProcessing}

Once the (physical or fictitious) surface density has been computed as the solution to the integral equation, all the post-processing facilities described in section \ref{secEqInt:quantite} are available. Note that computing the trace or normal derivative trace of the wavefield on the boundary of one of the disk depends on the integral representation of the scattered field, and generally only  implies a linear combination of the four boundary integral operators.

The post-processing functions are defined in the subdirectory \texttt{mudiff/PostProcessing/}. The function \texttt{PlotCircles}
allows to display the geometrical configuration given by the collection of disks. Functions related to the near-field 
are given by \texttt{ExternalPotential} and \texttt{InternalPotential} if one wants to compute the solution at a point of the domain or on a whole grid,
from the exterior or interior of the scatterers, respectively. In addition, far-fields can be obtained by applying the \texttt{FarField} function. For
 the Radar Cross Section, the $\mu$-diff function is called \texttt{RCS}. Each of these functions needs the integral representation of the scattered field.
To help the user, each function has an interface function for the single- and the double-layer potential only (e.g. \texttt{ExternalSingleLayerPotential}, \texttt{FarFieldSingleLayer}, \ldots). Even if the far-field is efficiently computed, the user should be aware that the computation of the volume potentials on a huge
discrete grid can need more time than assembling and solving the linear system. 

The reader can find in the example subdirectory \texttt{mudiff/Examples/Benchmark} many examples of manipulation of the code
in the files \texttt{BenchmarkDirichlet} (sound-soft scattering) and \texttt{BenchmarkNeumann} (sound-hard scattering).
An effort has been made to show all the possible combinations of operators available in $\mu$-diff, trying to use the main functions.
The user can clearly  play with the parameters sets, the only limit being given by the memory of the computer used.
The notation concerning the integral equations are related to the present paper (EFIE, MFIE, CFIE, BWIE).

\section{Numerical examples with $\mu$-diff}\label{sectionnumericalexamples}
\subsection{Example I: scattering by randomly distributed  sound-soft or sound-hard circular cylinders}\label{ExampleSimple}

To  show an example of problem solved by $\mu$-diff, we consider that we use the EFIE to solve the scattering problem
by a collection of sound-soft or sound-hard
 randomly distributed scatterers. The corresponding script (\texttt{BenchmarkDN}) for simulating the results of this section is available
  in  the examples directory. 

\begin{figure}
\centering
\subfigure[Cluster of $M=360$ disks]{\label{fig:scatO}\begin{tikzpicture}
 \pgftext{\includegraphics[width=0.45\textwidth]{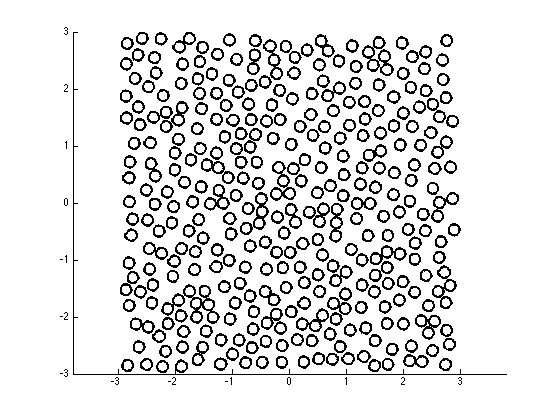}\label{configurationgeometry}}
 \draw (10pt,-65pt) node[below]{$x_{1}$} ;
 \draw (-90pt,10pt) node[below]{$x_{2}$} ;
\end{tikzpicture}}\quad\subfigure[RCS for the sound-soft and sound-hard cases]{\label{fig:scatRCS}\begin{tikzpicture}
 \pgftext{\includegraphics[width=0.45\textwidth]{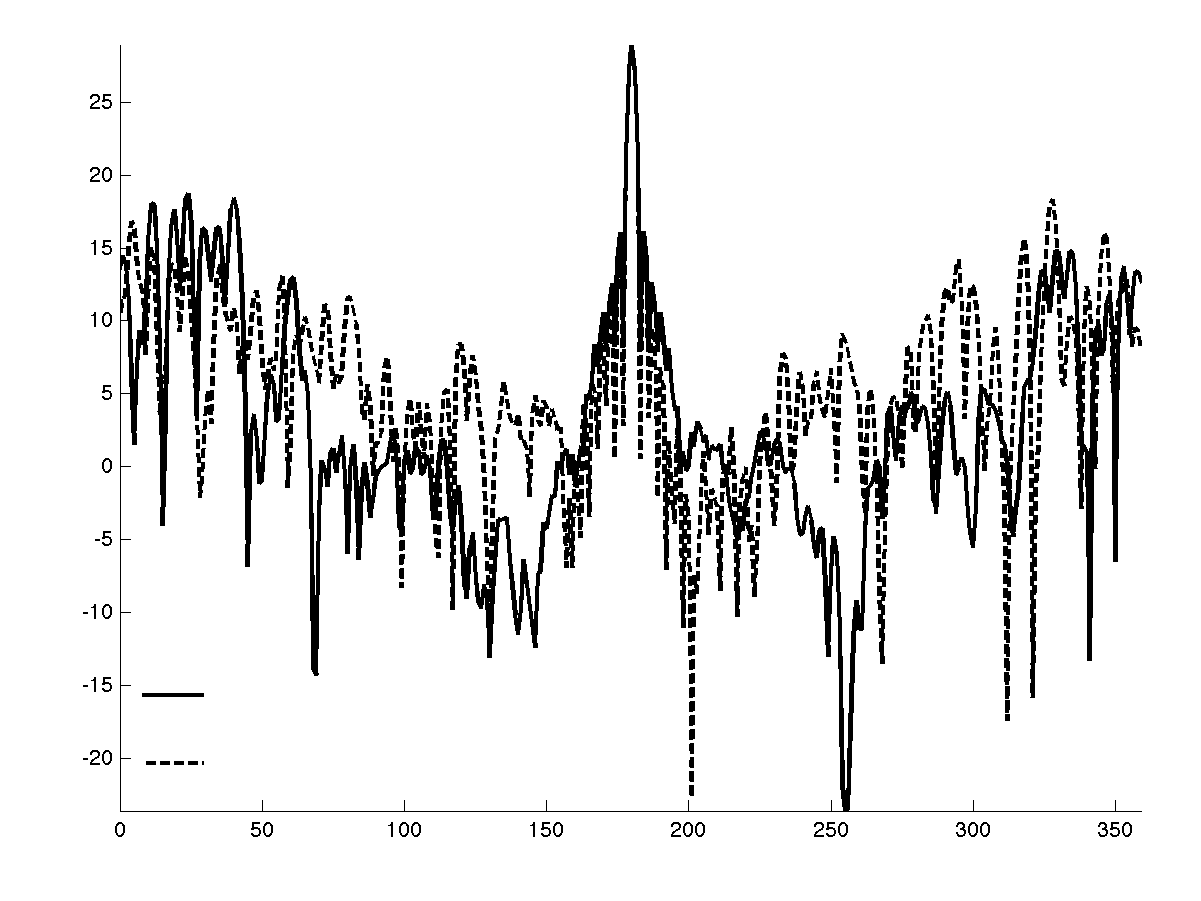}}
 \draw (10pt,-67pt) node[below]{\small Scattering angle $\theta$ (deg.)} ;
  \node[label={[label distance=0.5cm,text depth=-1ex,rotate=90]left:\small RCS (dB)}] at (-100pt,45pt) {};
 \draw (-49pt,-35pt) node[below]{\footnotesize Dirichlet} ;
 \draw (-47pt,-47pt) node[below]{\footnotesize Neumann} ;
\end{tikzpicture}}
\subfigure[Sound-soft problem: $|\ut|$]{\label{fig:scatD}\begin{tikzpicture}
 \pgftext{\includegraphics[width=0.48\textwidth]{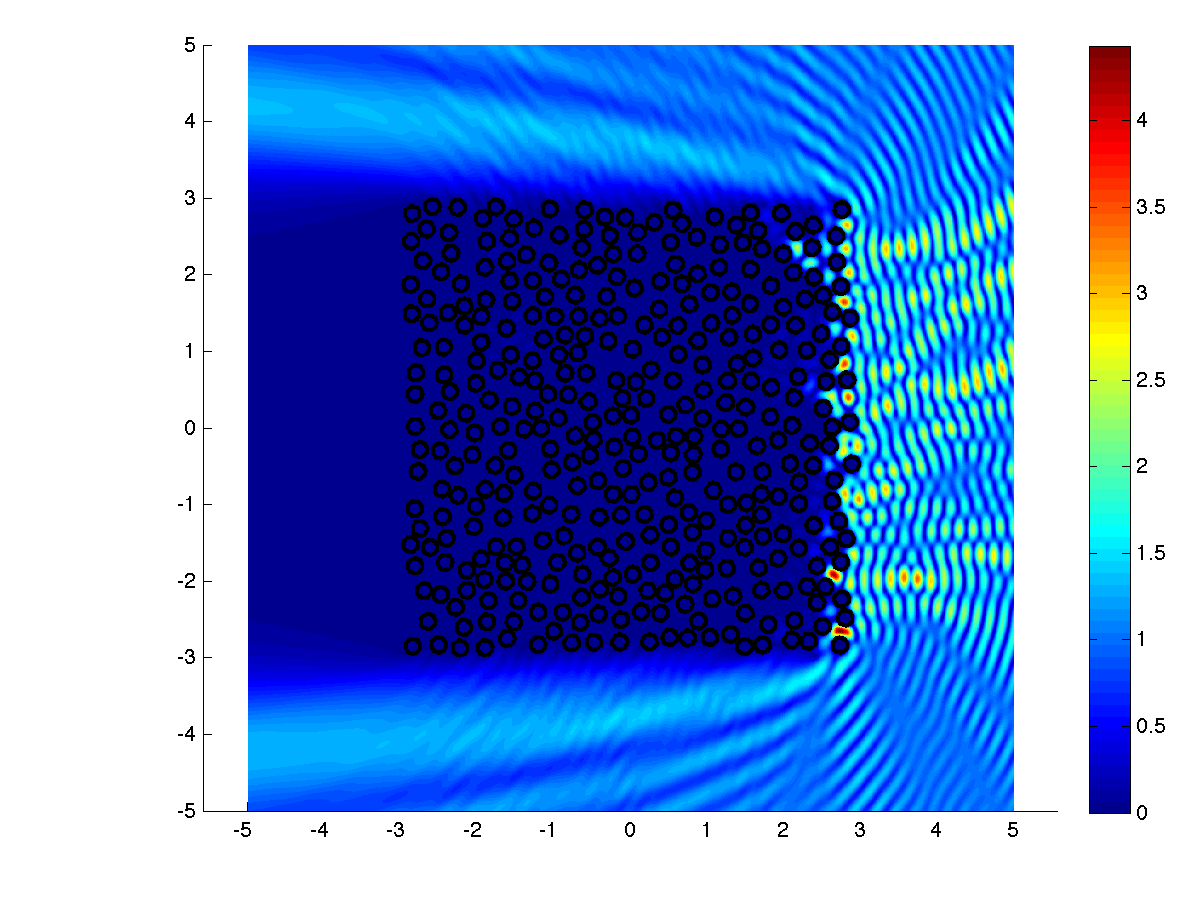}}
 \draw (10pt,-75pt) node[below]{\small $x_{1}$} ;
 \draw (-85pt,10pt) node[below]{\small $x_{2}$} ;
\end{tikzpicture}}\quad\subfigure[Sound-soft problem: $|u|$]{\label{fig:scatDs}\begin{tikzpicture}
 \pgftext{\includegraphics[width=0.48\textwidth]{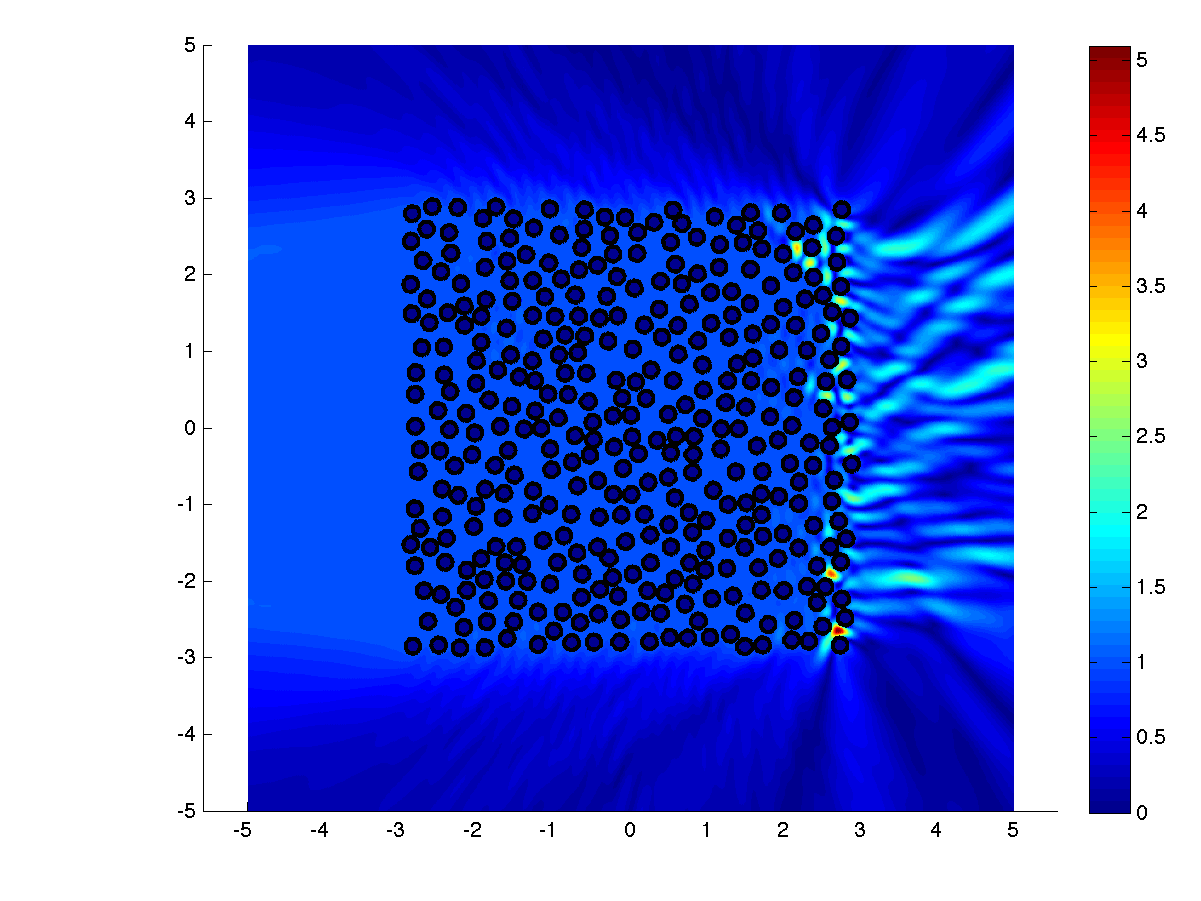}}
 \draw (10pt,-75pt) node[below]{\small $x_{1}$} ;
 \draw (-85pt,10pt) node[below]{\small $x_{2}$} ;
\end{tikzpicture}}
\subfigure[Sound-hard problem: $|\ut|$]{\label{fig:scatN}\begin{tikzpicture}
 \pgftext{\includegraphics[width=0.48\textwidth]{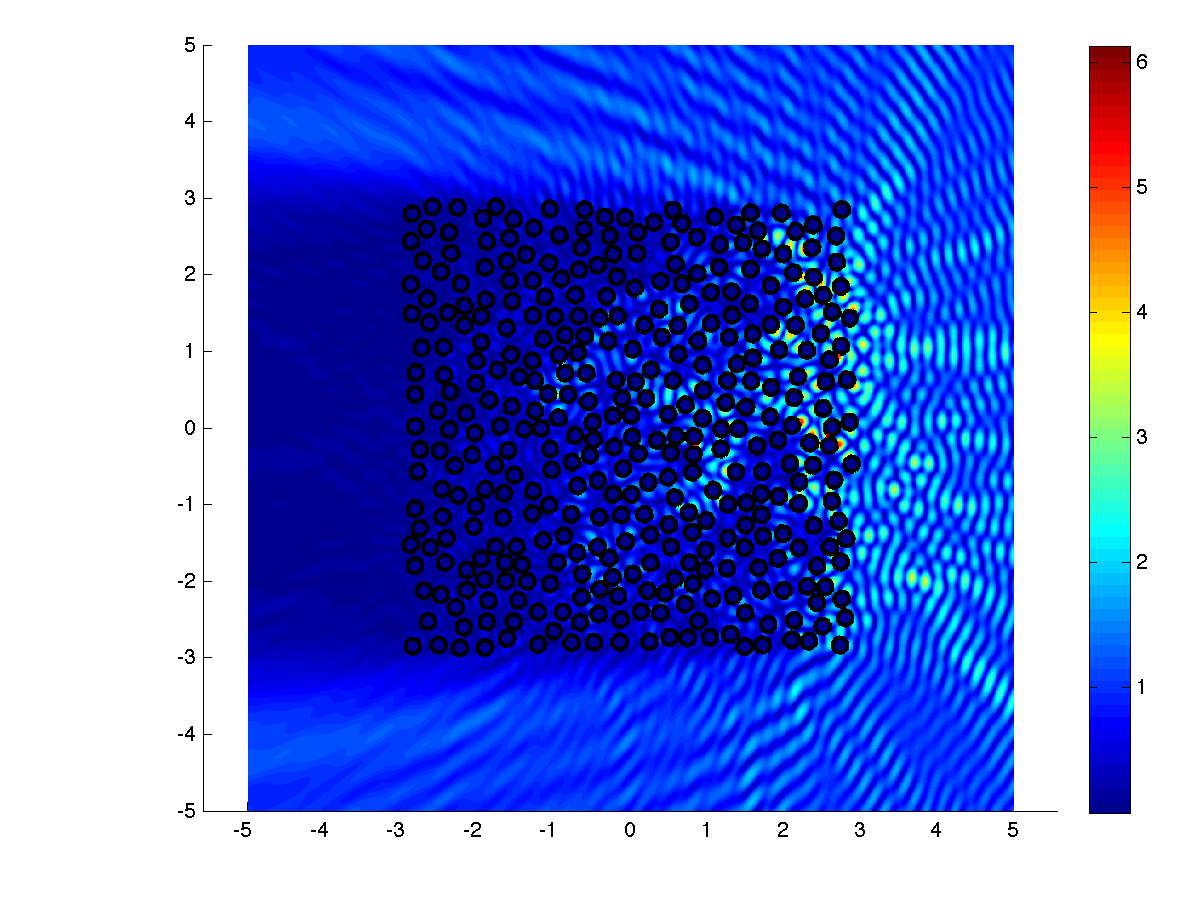}}
 \draw (10pt,-75pt) node[below]{$x_{1}$} ;
 \draw (-85pt,10pt) node[below]{$x_{2}$} ;
\end{tikzpicture}}\quad\subfigure[Sound-hard problem: $|u|$]{\label{fig:scatNs}\begin{tikzpicture}
 \pgftext{\includegraphics[width=0.48\textwidth]{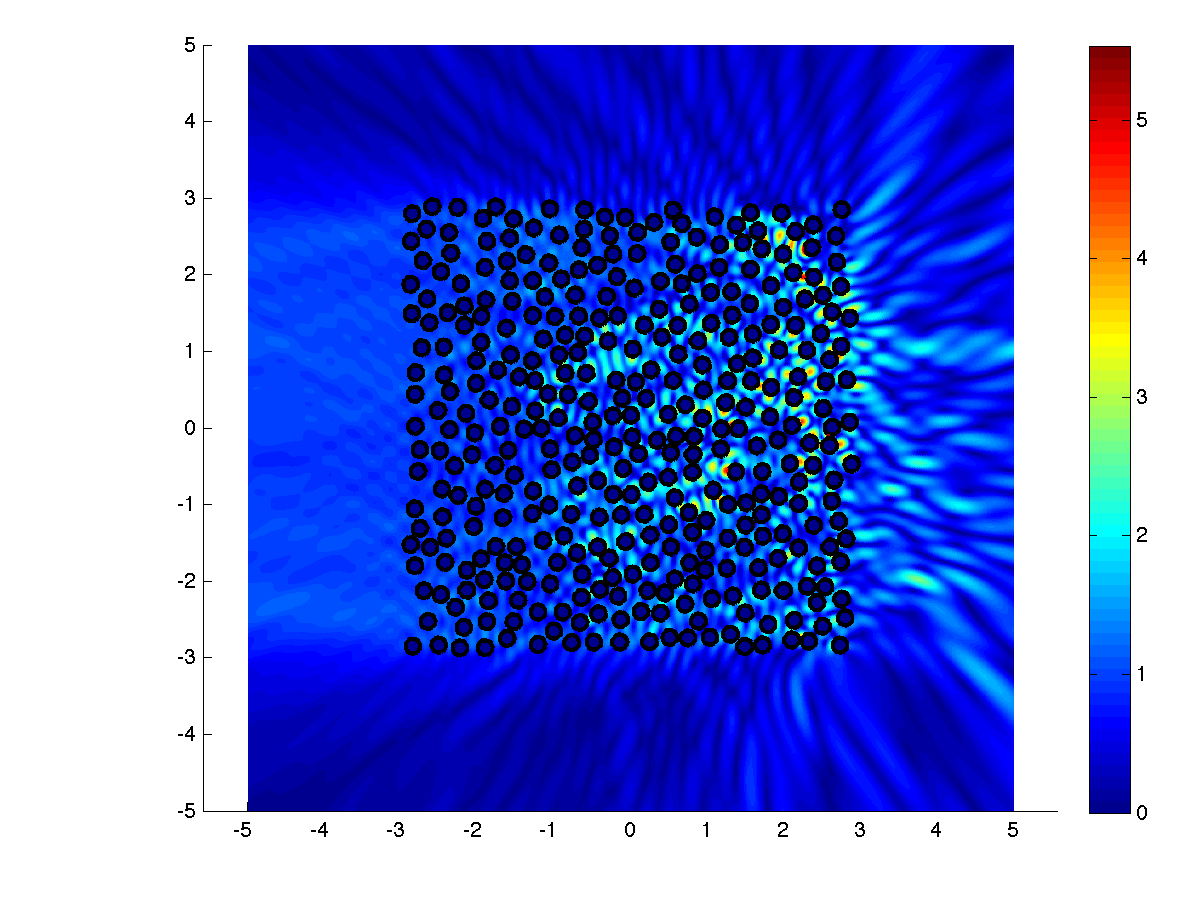}}
 \draw (10pt,-75pt) node[below]{$x_{1}$} ;
 \draw (-85pt,10pt) node[below]{$x_{2}$} ;
\end{tikzpicture}}
\caption{Multiple scattering of an incident plane wave ($k=6\pi$ and $\beta=\pi$ (rad.))
 by $M = 360$ sound-soft/sound-hard obstacles, randomly distributed in $[-3;3]^{2}$. }
\label{fig:scatO}
\end{figure}
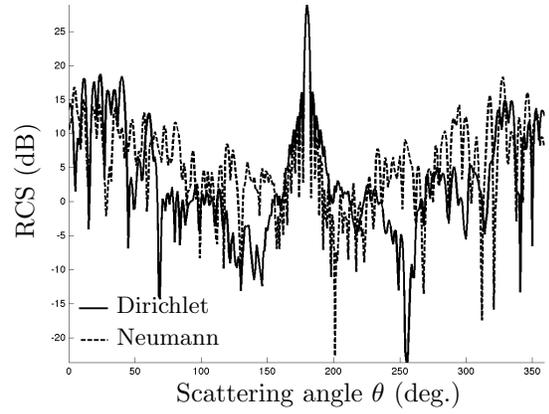
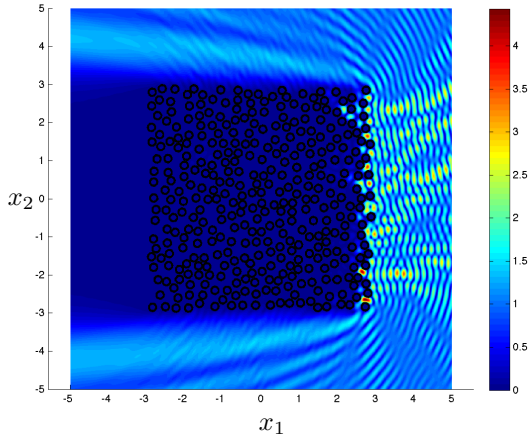
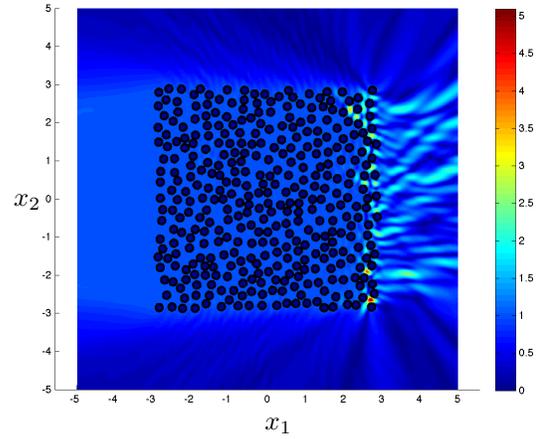
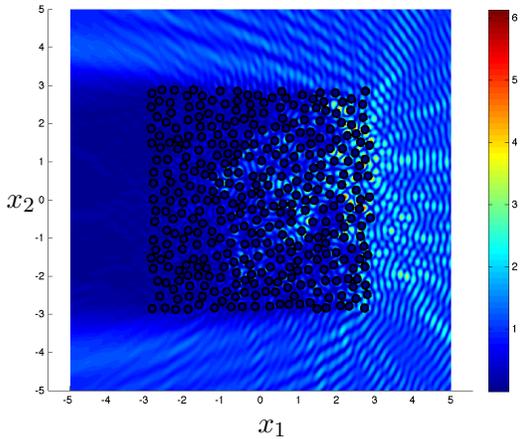
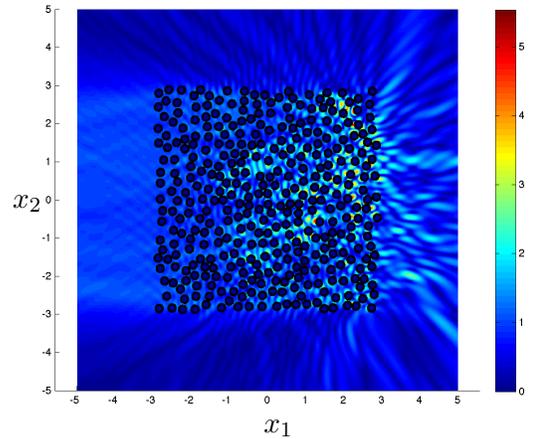

We consider a plane wave (for a wavenumber $k=6 \pi$ and an incidence angle $\beta=\pi$ (rad.)) that scatterers on a collection
of $M=360$ circular cylinders (see figure \ref{configurationgeometry}). These disks are randomly distributed in a square computational domain
 $[-3;3]^{2}$. In addition, their radii are
 such that $a_{\textrm{min}}:= 10^{-1} \leqslant a_{p}\leqslant a_{\textrm{max}} := 1.5\times 10^{-1}$, the minimal distance between the disks is $d_{\textrm{min}}:=0.01 \times a_{\textrm{min}}$. The number of modes  is fixed by the formula (\ref{eq:Np}), taken from (21) in \cite{AntChnRam08}.
The trace and normal derivative trace of the incident plane wave are then defined to build the right-hand sides of the EFIE.
We report in figure \ref{fig:scatRCS} the RCS for the sound-soft and sound-hard acoustic problems. 
These pictures show that the far-fields have some very different
structures. In addition, the amplitudes of total and scattered wavefields are displayed on figures \ref{fig:scatD}-\ref{fig:scatDs} for
the sound-soft problem and figures \ref{fig:scatN}-\ref{fig:scatNs} for
the sound-hard problem. We consider a larger computational domain to show the wavefield behavior
both inside and outside
the cluster of circular cylinders. We observe in particular that there is almost no penetration of the incident field
in the sound-soft case while scattering arises deeply in the sound-hard cluster.

\subsection{Example II: multiple scattering by a cluster of homogeneous 
penetrable obstacles}\label{sec:penetrable}

Extending the previous example, the script \texttt{BenchmarkPenetrable} solves the transmission problem with penetrable obstacles. The wavenumber $k$ is now piecewise constant with value $k^+$ outside the obstacles and $\kint$ inside. The scattered field $\ups$ and the
transmitted wavefield $\um$ are then the solution to the following transmission boundary-value problem
\begin{equation}
\label{eq:PenetrableSyst}
\left\{\begin{array}{r c l l}
\Delta \um + (\kint)^2\um &=& 0, & \text{in }\Omegam,\\
\Delta \ups + (k^+)^2\ups &=& 0, & \text{in }\Omegaps,\\
\ups-\um &=& -\uinc, & \text{on }\Gamma,\\
\dn\ups-\dn\um &=& -\dn\uinc, & \text{on }\Gamma,\\
\multicolumn{4}{l}{\displaystyle {\lim_{||\mathbf{x}||\to + \infty} ||\mathbf{x}||^{1/2}\left(\nabla \ups\cdot \frac{\mathbf{x}}{||\mathbf{x}||}-i k u\right)=0.} }
\end{array}\right.
\end{equation}
The total (physical) field $\ut$ is given by $\ut = \ups + \uinc$ outside and by $\ut=\um$ inside the obstacles. To solve this 
problem through an integral equation, we consider a single-layer representation of the wavefields
 $\ups$ and $\um$ 
\begin{equation}\label{eq:penetrableIF}
\ups = \Lop^+\rhops\quad\text{ and }\quad \um = \Lop^-\rhomi,
\end{equation}
where $\Lop^+$ (resp. $\Lop^-$) is the single-layer operator with wavenumber $k^+$ (respectively $\kint$).
 The pair of unknowns $(\rhops,\rhomi)$ is then  the solution to the following integral equation
$$
\left(\begin{array}{c c}
L^+ & -L^- \\
\displaystyle -\frac{I}{2} + N^+ & \displaystyle \left(\frac{I}{2}+N^-\right)
\end{array}
\right)
\left(\begin{array}{c}
\rhops\\\rhomi
\end{array}
\right)
=
\left(\begin{array}{c}
-\uinc\\-\dn\uinc
\end{array}
\right).
$$
The plus or minus superscripts in $L$ and $N$ refers to as the exterior wavenumbers $k^+$ or $\kint$.
 Like for the sound-soft and sound-hard  scattering problems, the far-field 
  and the quantities $\ups$ and $\um$ can be computed, thanks to their respective 
  single-layer representation (\ref{eq:penetrableIF}). 
Let us remark that the present problem also arises for electromagnetic wave scattering by dielectric obstacles. 
The wavenumbers are then given by $k^+=\omega\sqrt{\e_0\mu_0}$ and $\kintp=\omega\sqrt{\e_p\mu_p}$, where $\omega$ is the pulsation
of the wave and $(\e_0,\mu_0)$ (respectively $(\e_p,\mu_p)$) are respectively the electric permittivity and electromagnetic permeability in the vacuum 
(respectively in the obstacle $\Omega_p$). The equation (\ref{eq:PenetrableSyst}) remains the same except for the fourth line which
is now: $\dn\ups-\mu\dn\um = -\dn\uinc$, on $\Gamma$, where $\mu|_{\Omega_p} =\mu_p$. As a consequence,
 the integral equation in only changed by multiplying $(I/2+N^-)$ by the parameter $\mu$.

\begin{figure}
\centering
\subfigure[RCS for the penetrable case]{\label{fig:penetrableRCS}\begin{tikzpicture}
 \pgftext{\includegraphics[width=0.45\textwidth]{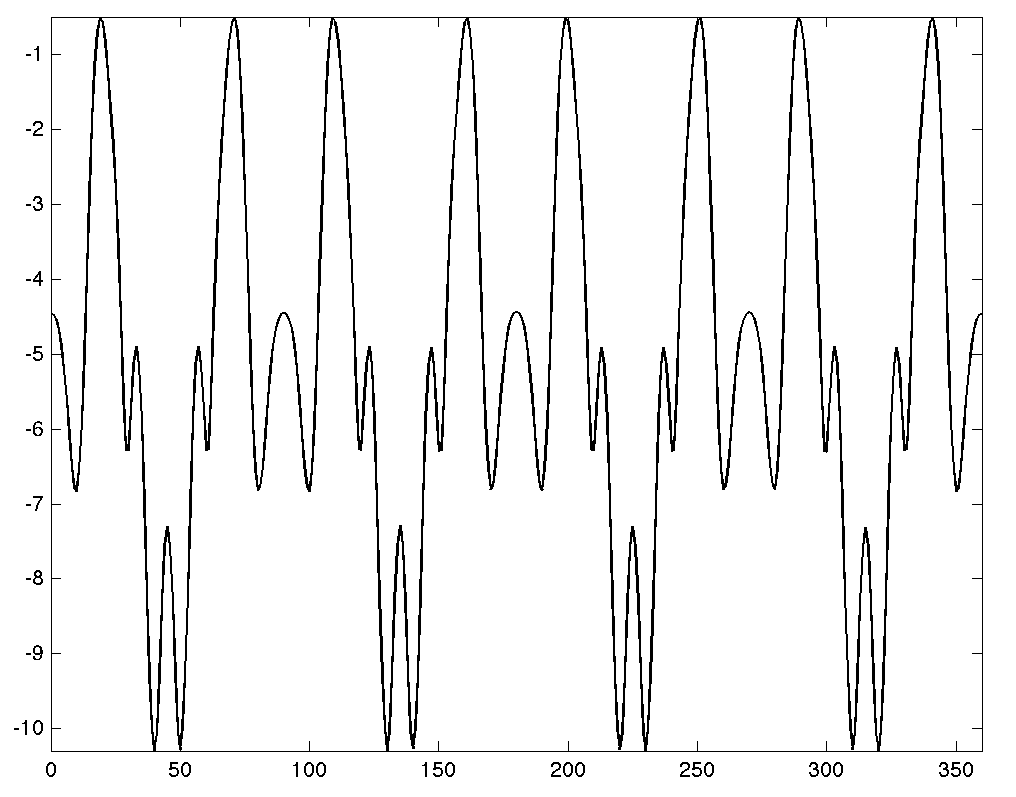}}
 \draw (-5pt,-75pt) node[below]{{\small Scattering angle $\theta$ (deg.)}} ;
  \node[label={[label distance=0.5cm,text depth=-1ex,rotate=90]left:{\small{RCS (dB)}}}] at (-110pt,45pt) {};
\end{tikzpicture}}\subfigure[Penetrable problem: $\left|\ut\right|$]{\label{fig:penetrableABS}\begin{tikzpicture}
 \pgftext{\includegraphics[width=0.45\textwidth]{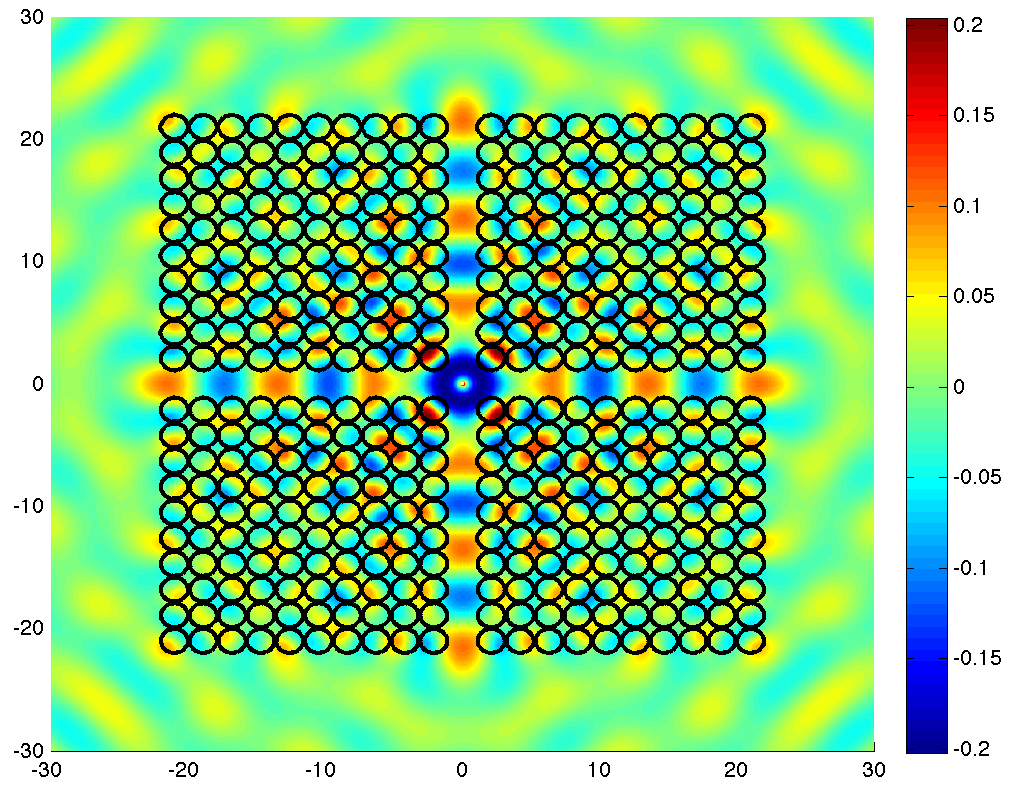}}
 \draw (-5pt,-78pt) node[below]{\small $x_{1}$} ;
 \draw (-110pt,10pt) node[below]{\small $x_{2}$} ;
\end{tikzpicture}
}\newline\subfigure[Penetrable problem: $\Re(\ut)$]{\label{fig:penetrableRE}\begin{tikzpicture}
 \pgftext{\includegraphics[width=0.45\textwidth]{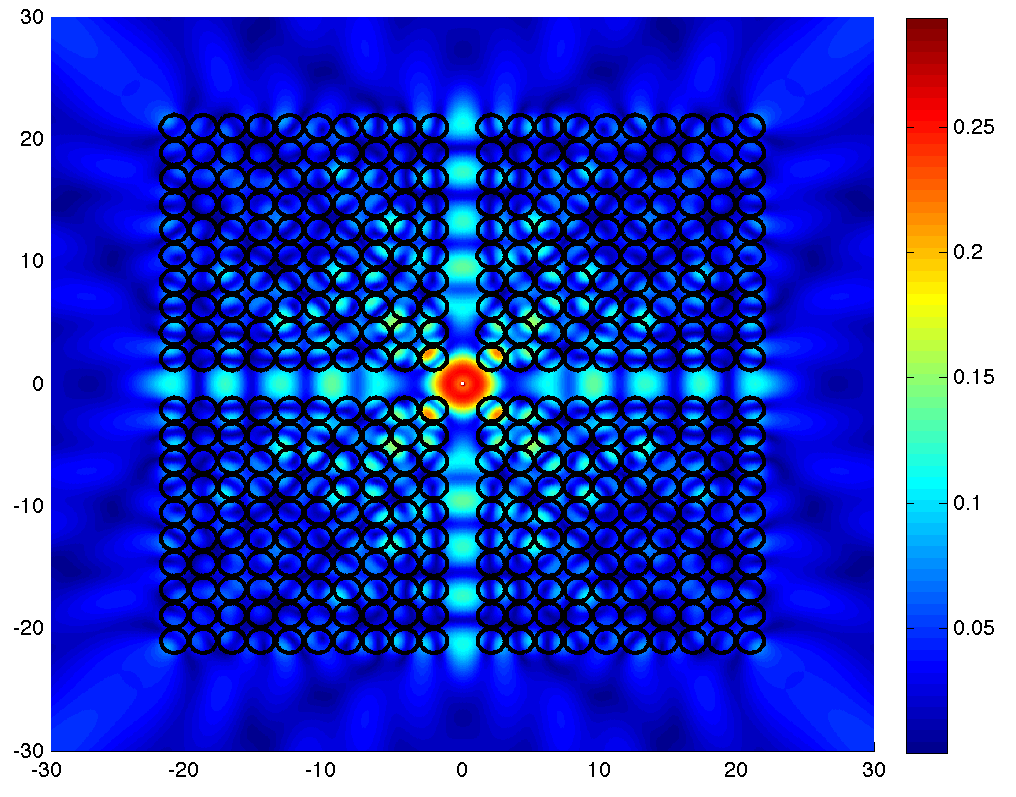}}
 \draw (-5pt,-80pt) node[below]{\small $x_{1}$} ;
 \draw (-110pt,10pt) node[below]{\small $x_{2}$} ;
\end{tikzpicture}
}\subfigure[Penetrable problem: $\Im(\ut)$]{\label{fig:penetrableIM}\begin{tikzpicture}
 \pgftext{\includegraphics[width=0.45\textwidth]{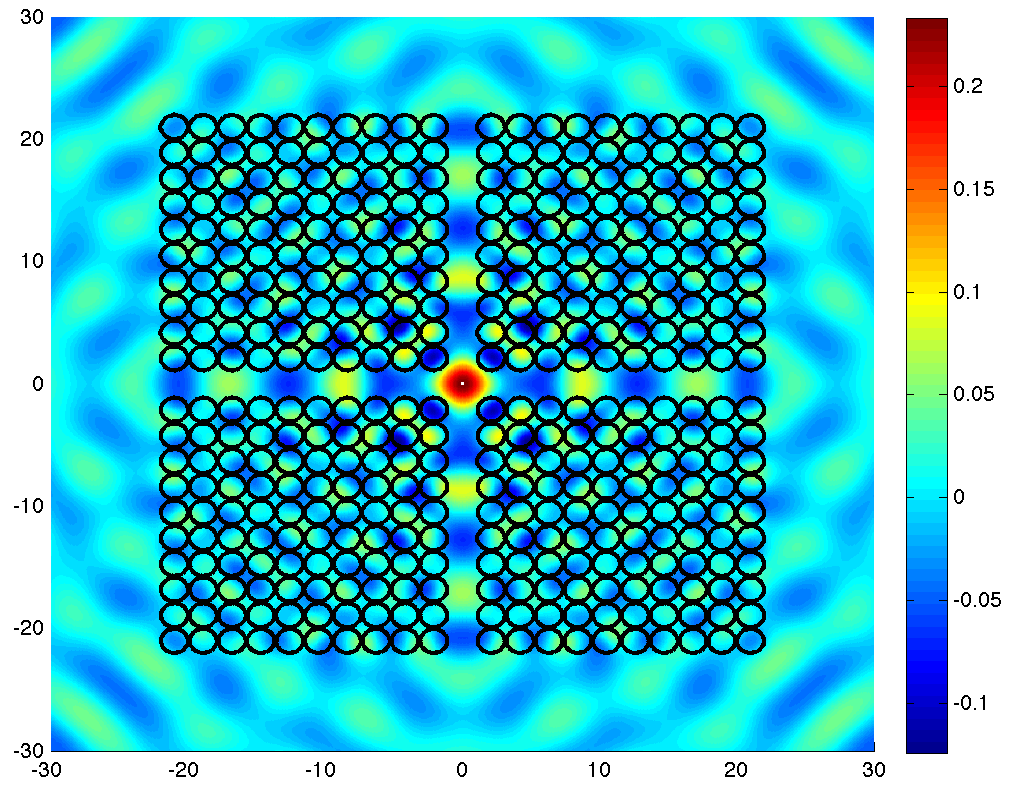}}
 \draw (-5pt,-80pt) node[below]{\small $x_{1}$} ;
 \draw (-110pt,10pt) node[below]{\small $x_{2}$} ;
\end{tikzpicture}
}
\caption{Scattering of a point source (located at the origin) by
  a collection of $M=400$ penetrable unit disks (interior wavenumber $k^-=2k$, with the exterior wavenumber $k=1$).}
\label{fig:penetrable}
\end{figure}
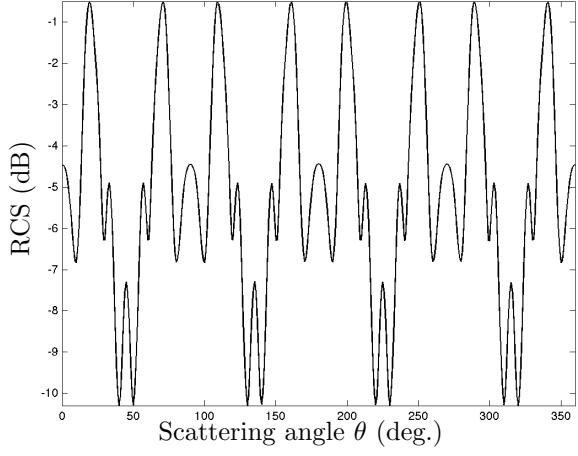
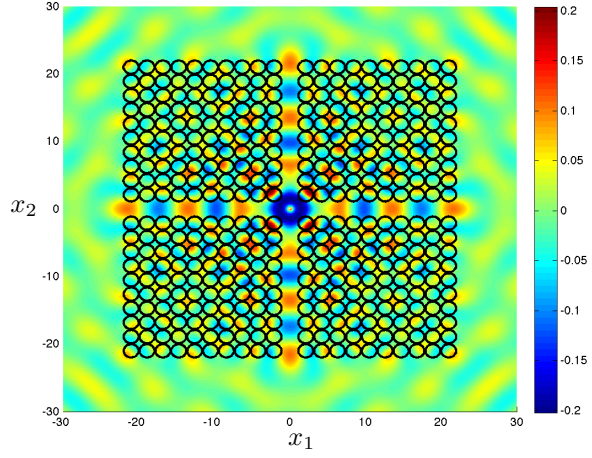
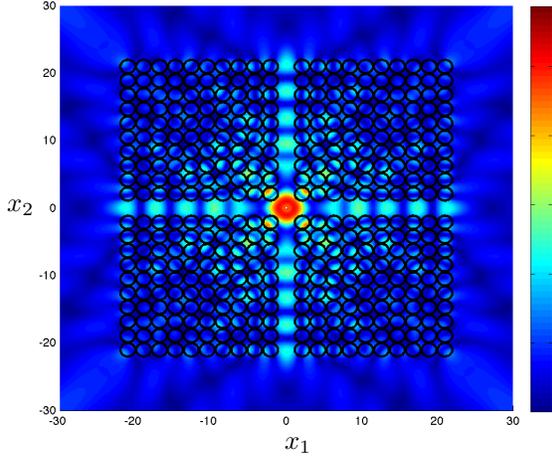
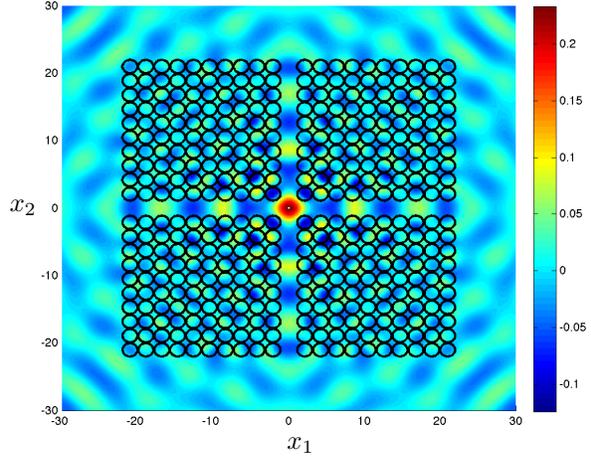

A numerical example solved by $\mu$-diff is shown in figures \ref{fig:penetrableRCS}-\ref{fig:penetrableIM} for  $M=400$ unit penetrable unitary disks placed as a rectangular lattice centered on $(0,0)$, which is also the location of a point source emitting a wave. The middle row and column, corresponding to centers with $0$ abscissa and $0$ ordinate respectively, have been removed. The whole geometry has been built thanks to the pre-processing $\mu$-diff functions  \texttt{RectangularLattice} and \texttt{RemoveDisk}. This last function  deletes easily some  disks in a geometrical configuration if they are not needed.
 The exterior wavenumber is set to $k^+=1$ and the wavenumber $k^-$ inside the obstacles is equal to $k^-=2k^+=2$. We report the RCS, as well as the amplitude, real and imaginary parts of the total field $\ut$. Of course, more scatterers, higher frequencies and complex-valued wavenumbers could be chosen when launching a simulation with $\mu$-diff.

\subsection{Example III: a more advanced application in time reversal}\label{sectionDORT}

Finally, a last numerical example related to the use of $\mu$-diff concerns an inverse scattering problem.
Time reversal is a technique based on the reversibility property of the wave equation in a non dissipative medium to send back a signal in the original medium and on the source that first emitted it. The goal is to get informations about the medium. 
Time reversal methods do not provide  a full characterization of the medium but lead to
 some useful informations about the presence of  failures or obstacles in the medium.
Physical time reversal experiments  are possible since the pioneering developments of the Time Reversal Mirror (TRM) by Fink and his team \cite{fin08}. These devices are composed by numerous cells that can play alternatively the role of emitters or receivers. A typical time reversal experiment can be described as follows: a point source emits a wave in the medium, the mirror measures it, time-reverses it and sends it back to the medium. The resulting back-propagated wave is expected to focus on the source both in space and time, with a resolution depending on various parameters such as the size, the position or the distance of the mirror
 to the source, the medium,\ldots (without being exhaustive, we refer for instance to \cite{ BorPapTso03,Fin06}, the literature on this topic being huge). 

Based on this idea, the DORT method (french acronym for ``Decomposition of the Time Reversal Operator''), developed by Prada and Fink \cite{FinPra94, Pra97}, aims to detect and locate non-emitting objects. In fact, this technique go further than the pure detection since it also generates waves that focus selectively on the obstacles
that are supposed to be small and distant enough from
 each other. The fascinating applications of the DORT are numerous. Let us mention among others the imaging and
 the destruction of kidney stones  \cite{MasNacWaa97,ThoWuFin96} or subsurface imaging \cite{BorPapTso11} by using the DORT as a filter. The method is based on the iteration of the following cycle: first the TRM emits a wave toward the obstacles, generating a scattered wave which is then measured by the TRM and time-reversed. This cycle - ''emission, reception and time-reversal'' - is then repeated again by sending back the time-reversed measurements. After many iterations, it appears that the back propagated wave focuses on the most reflecting obstacle. To detect and focus waves on the other obstacles, the DORT method consists in 1) building the so-called Time Reversal Operator, designated by $\mathcal{T}$ here and
  defined by two cycles ''emission-reception-time reversal'', and 2)
 study its spectral properties. Indeed, when the obstacles are small and sufficiently far to each other, this  operator has as much significant eigenvalues as
  the number of obstacles, and moreover, the associated eigenfunctions can be used to generate waves that focus selectively on the obstacles. This has been proved mathematically in the far-field context for sound-hard acoustic scattering in \cite{HazRam04} and studied  numerically by
 using an earlier basic version of the $\mu$-diff toolbox in \cite{ThierryThesis}. These results have also been extended to other types of waves such as the dielectric cases in \cite{BurMinRam13} for which numerical simulations have been performed. 
  
 For the sake of simplicity, we only present here
  the acoustic far-field case, even if the scripts for the two cases are available in  the \texttt{Examples/TimeReversal/FarField} directory of the 
  current $\mu$-diff toolbox. 
For this case, the time reversal mirror is placed at infinity and totally surrounds the obstacles.  In particular, this implies that the TRM sends a linear combination of plane waves, called Herglotz waves, and measures the scattered far-field. More precisely, an Herglotz wave $u_I$ with parameter $f$ is given by $$u_I(x_{1},x_{2})
 = \int_{0}^{2\pi} f(\alpha) e^{ik(x_{1}\cos(\alpha) +x_{2}\sin(\alpha))}\;{\rm d}\alpha.$$ 
 Let us denote by $\mathcal{F}f$ the far-field generated by an Herglotz wavefield of parameter $f$.
 Then, it can be proved \cite{HazRam04} that the TRO is given by: $\mathcal{T} = \mathcal{F}^*\mathcal{F}$, where $\mathcal{F}^*$ is the adjoint operator of $\mathcal{F}$.
 An eigenfunction $g$ of $\mathcal{T}$ can then be used as a parameter of an Herglotz wavefunction to generate a wave   focusing on the
  obstacles if its associated eigenvalue is significantly large.
In the discrete context, building the matrix $\mathbb{T}$ associated with the operator $\mathcal{T}$ can be done as follows. First, the TRM is discretized by
using $N_{\alpha}$ points or angles $\alpha_{j}$, $j=1,...,N_{\alpha}$ (note that, if a point emits an incident wave with angle $\alpha$, then
the TRM measures the far-field
 in the opposite direction $\alpha+\pi$). A discrete Herglotz wave emitted by the mirror is then  
$$u_I(x_{1},x_{2}) = \sum_{j=1}^{N_\alpha} h_\alpha f_j e^{ik(x_{1}\cos(\alpha_j) +x_{2}\sin(\alpha_j))},$$
where $f_j = f(\alpha_j)$ and $h_\alpha$ is the discretization step. The algorithm to obtain the time reversal matrix $\mathbb{T}$ is then : for every angle $\alpha_j$, the scattered field is computed and the associated far-field is stored in a matrix $\mathbb{F}(:,j)$ of size $N_{\alpha}\times N_{\alpha}$. Once
$\mathbb{F}$ has been computed,  the matrix $\mathbb{T}$ is obtained by the relation: $T=\overline{\mathbb{F}}^T \mathbb{F}$.  

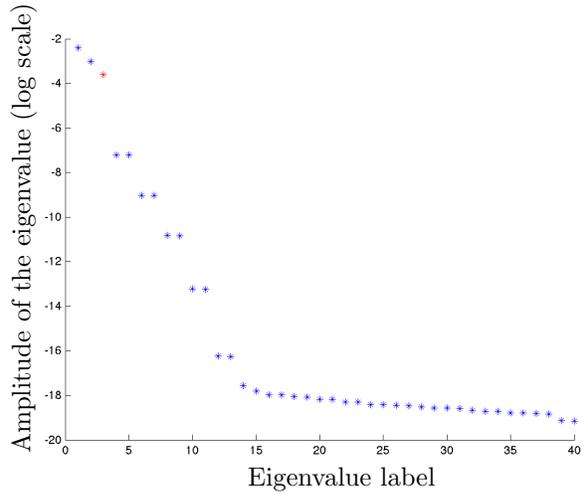
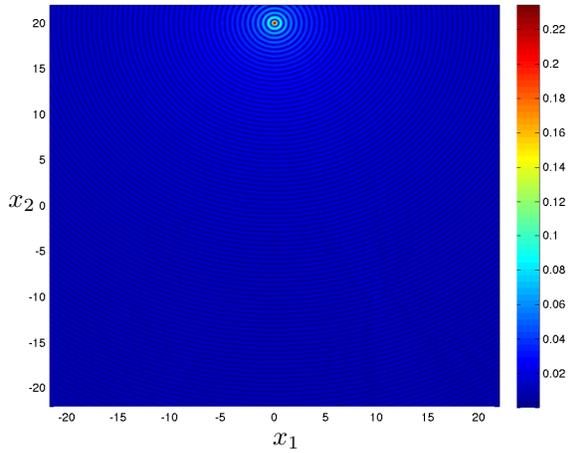
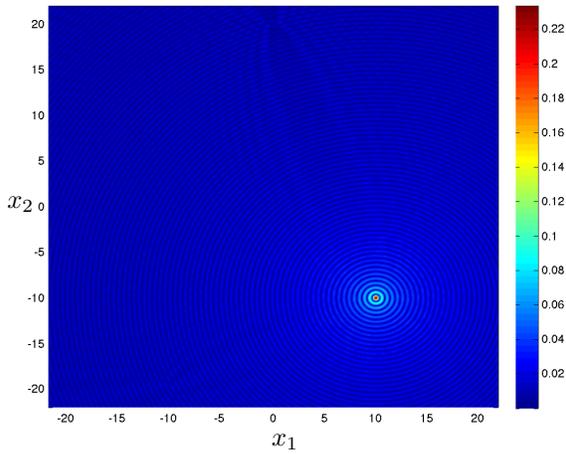
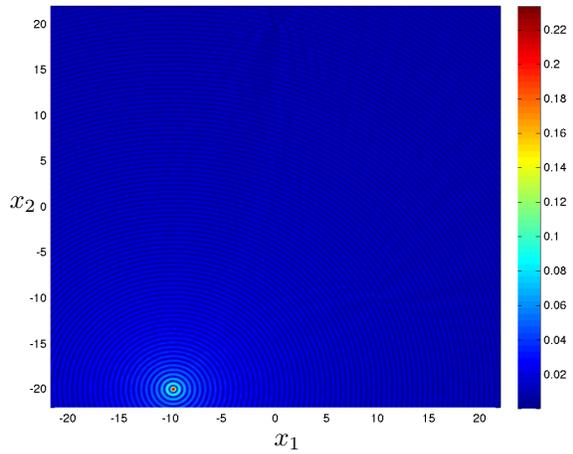
\begin{figure}
\centering
\subfigure[$40$ largest eigenvalues of $T$ ($\log$ scale)]{\label{fig:DORTE}\begin{tikzpicture}
 \pgftext{\includegraphics[width=0.45\textwidth]{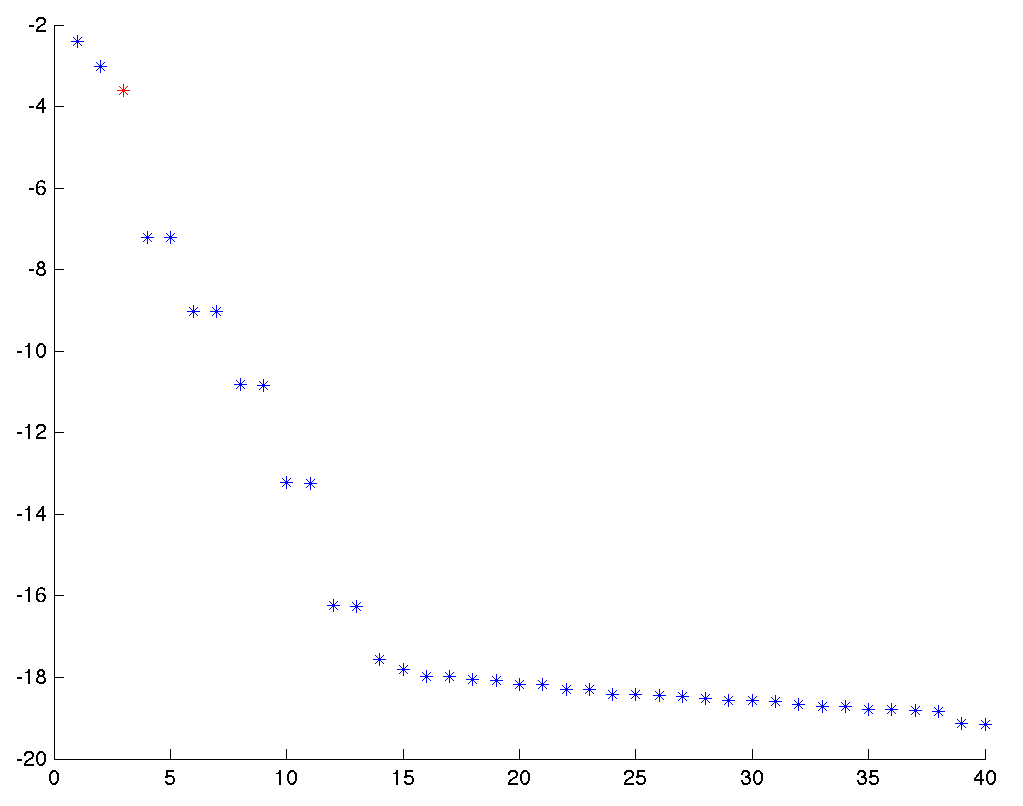}}
 \draw (10pt,-80pt) node[below]{\small Eigenvalue label} ;
  \node[label={[label distance=0.5cm,text depth=-1ex,rotate=90]left:\small Amplitude of the eigenvalue (log scale)}] at (-110pt,110pt) {};
\end{tikzpicture}
}\subfigure[Amplitude of the first Herglotz wavefunction]{\label{fig:DORTH1}\begin{tikzpicture}
 \pgftext{\includegraphics[width=0.45\textwidth]{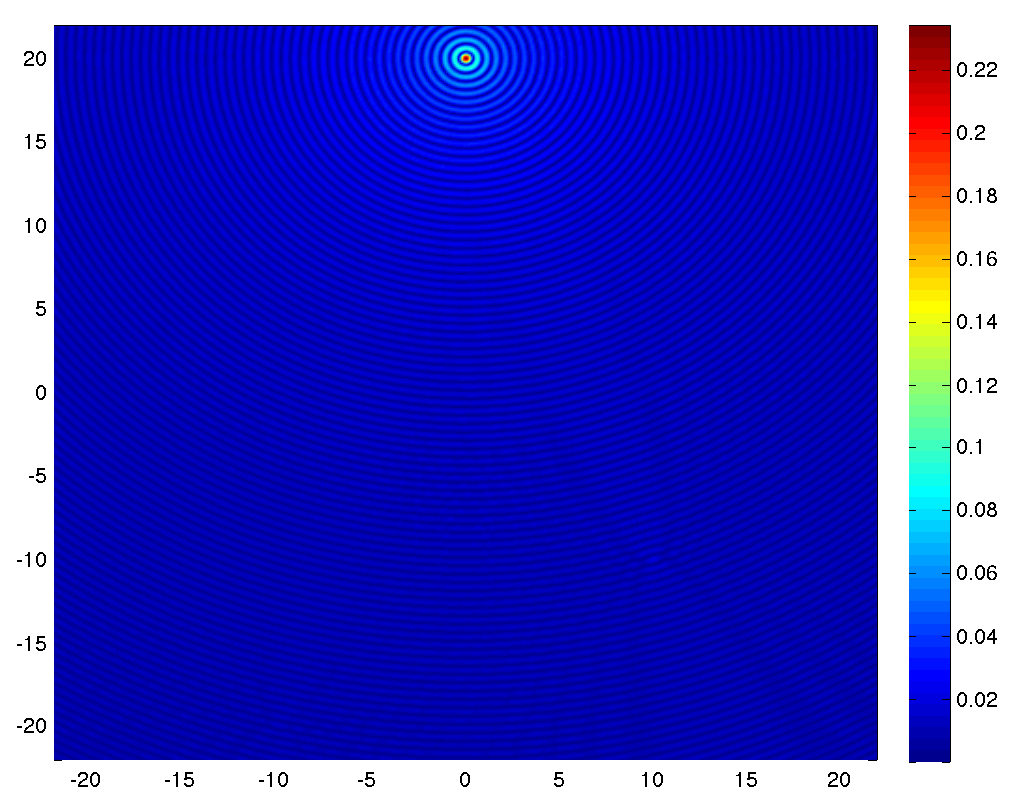}}
 \draw (-5pt,-80pt) node[below]{\small $x_{1}$} ;
 \draw (-105pt,10pt) node[below]{\small $x_{2}$} ;
\end{tikzpicture}
}\newline\subfigure[Amplitude of the second Herglotz wavefunction]{\label{fig:DORTH2}\begin{tikzpicture}
 \pgftext{\includegraphics[width=0.45\textwidth]{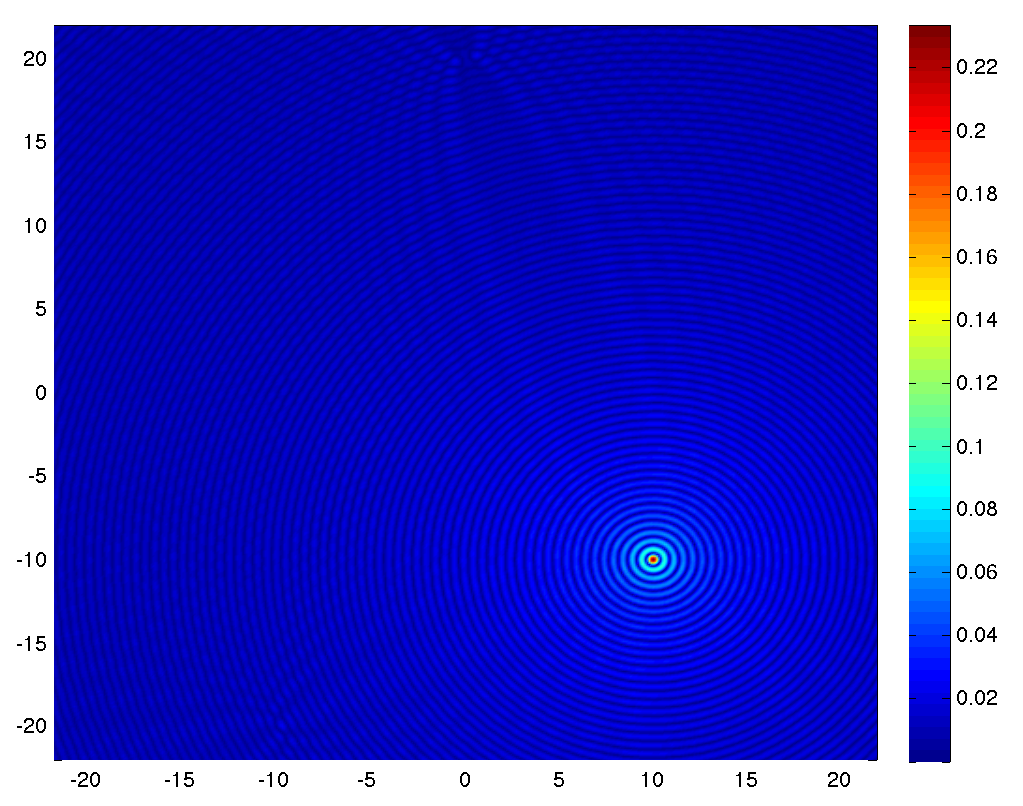}}
 \draw (-5pt,-80pt) node[below]{\small $x_{1}$} ;
 \draw (-105pt,10pt) node[below]{\small $x_{2}$} ;
\end{tikzpicture}
}\subfigure[Amplitude of the third Herglotz wavefunction]{\label{fig:DORTH3}\begin{tikzpicture}
 \pgftext{\includegraphics[width=0.45\textwidth]{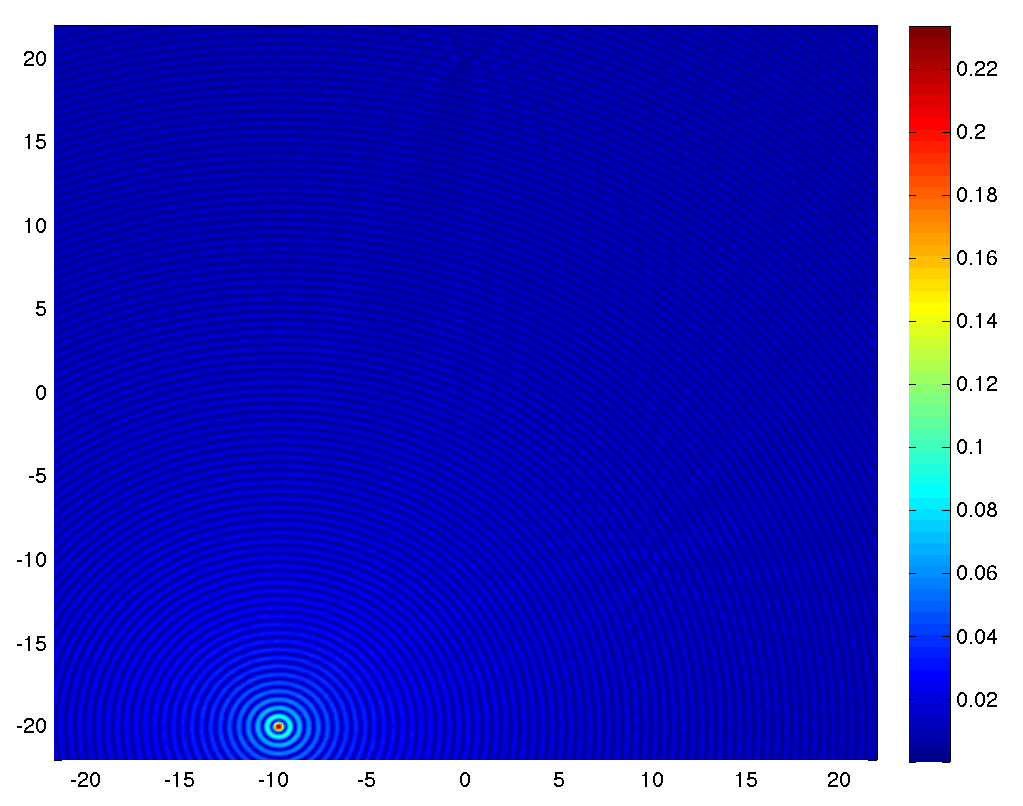}}
 \draw (-5pt,-80pt) node[below]{\small $x_{1}$} ;
 \draw (-105pt,10pt) node[below]{\small $x_{2}$} ;
\end{tikzpicture}
}
\caption{DORT: an example of numerical experiment  obtained by using $\mu$-diff.}
\label{fig:DORT}
\end{figure}

All the elementary operations described above can be easily coded by using $\mu$-diff  and the Matlab function \texttt{eigen} which
 provides  the eigenvalues and eigenvectors of $\mathbb{T}$. The Herglotz waves are computed thanks to the  function \texttt{HerglotzWave} available
  in the $\mu$-diff directory related to the examples. Finally, running the script \texttt{DORT\_Impenetrable.m} generates a DORT experiment.
  An example is given on figures \ref{fig:DORTE}-\ref{fig:DORTH3}.  We consider a medium with three penetrable circular scatterers, 
  with centers $[0,20]$, $[10,-10]$, $[-10,-20]$ and  respective radius    $0.02$, $0.01$, $0.005$.
  The  wavenumber is equal to  $k=2\pi$. 
   As shown on figure \ref{fig:DORTE}, the time reversal matrix $\mathbb{T}$ has three significant eigenvalues.
   We report on  figures \ref{fig:DORTH1}-\ref{fig:DORTH3} the amplitude of the  Herglotz wavefunctions associated with the three largest eigenvalues.
   We clearly observe that  they selectively focus on the obstacles, from the most to the less reflecting (or largest) one.

\section{Conclusion}\label{sectionConclusion}

This paper presented a new flexible, efficient and robust Matlab toolbox called $\mu$-diff\footnote{\url{http://mu-diff.math.cnrs.fr}}. This open source code
 is based on the theory of integral representations  for solving two-dimensional multiple scattering problems by many circular cylinders. 
 The spectral approximation method uses Fourier series expansion and efficient linear algebra algorithms in conjunction
 with optimized memory storage techniques for solving the finite-dimensional 
 approximate integral formulations. Pre- and post-processing facilities are included in $\mu$-diff (near- and far-fields representations,
 surface fields). All the features are described
 with enough details so that the user can directly solve complex problems related to physics or engineering applications. In addition, we provide some benchmark scripts
 that reproduce the simulations shown in this paper (direct and inverse scattering). The $\mu$-diff toolbox is developed in such a way that 
 a wide class  of multiple
 scattering problems by disks can be solved.

\medskip 

\textbf{Acknowledgments.}
This work has been  funded by the Institute of Scientific Research and Revival of Islamic Heritage at Umm Al-Qura University (project ID 43405027) 
and the French National Agency for Research (ANR)  (project MicroWave NT09 460489).

\bibliographystyle{plain}
\bibliography{Biblio_mudiff.bib}

\begin{thebibliography}{10}

\bibitem{AcostaOSRC}
S.~Acosta.
\newblock {On-surface radiation condition for multiple scattering of waves}.
\newblock {\em {Computer Methods in Applied Mechanics and Engineering}}, {in
  press}, {2014}.

\bibitem{AcostaVillamizar}
S.~Acosta and V.~Villamizar.
\newblock { Coupling of Dirichlet-to-Neumann boundary condition and finite
  difference methods in curvilinear coordinates for multiple scattering}.
\newblock {\em {J. Comput. Phys.}}, {229}({5498-5517}), {2010}.

\bibitem{AntChnRam08}
X.~Antoine, C.~Chniti, and K.~Ramdani.
\newblock On the numerical approximation of high-frequency acoustic multiple
  scattering problems by circular cylinders.
\newblock {\em J. Comput. Phys.}, 227(3):1754--1771, 2008.

\bibitem{AntoineDarbasQJMAM}
X.~Antoine and M.~Darbas.
\newblock Alternative integral equations for the iterative solution of acoustic
  scattering problems.
\newblock {\em Quaterly J. Mech. Appl. Math.}, 1(58):107--128, 2005.

\bibitem{AntoineDarbasM2AN}
X.~Antoine and M.~Darbas.
\newblock Generalized combined field integral equations for the iterative
  solution of the three-dimensional {H}elmholtz equation.
\newblock {\em M2AN Math. Model. Numer. Anal.}, 1(41):147--167, 2007.

\bibitem{AntoineDarbasBook}
X.~Antoine and M.~Darbas.
\newblock {\em Integral Equations and Iterative Schemes for Acoustic Scattering
  Problems}.
\newblock to appear, 2014.

\bibitem{AntGeuRam10}
X.~Antoine, C.~Geuzaine, and K.~Ramdani.
\newblock {\em Wave Propagation in Periodic Media - Analysis, Numerical
  Techniques and Practical Applications}, volume~1, chapter Computational
  Methods for Multiple Scattering at High Frequency with Applications to
  Periodic Structures Calculations, pages 73--107.
\newblock Progress in Computational Physics, 2010.

\bibitem{JACT}
X.~Antoine, K.~Ramdani, and B.~Thierry.
\newblock Wide frequency band numerical approaches for multiple scattering
  problems by disks.
\newblock {\em J. Algorithms Comput. Technol.}, 6(2):241--259, 2012.

\bibitem{ISI:000253549800023}
S.~Bidault, F.J.G. de~Abajo, and A.~Polman.
\newblock {Plasmon-based nanolenses assembled on a well-defined DNA template}.
\newblock {\em {Journal of the American Chemical Society}}, {130}({9}):{2750+},
  {2008}.

\bibitem{BorPapTso03}
L.~Borcea, G.~Papanicolaou, and C.~Tsogka.
\newblock A resolution study for imaging and time reversal in random media.
\newblock In {\em Inverse problems: theory and applications ({C}ortona/{P}isa,
  2002)}, volume 333 of {\em Contemp. Math.}, pages 63--77. Amer. Math. Soc.,
  Providence, RI, 2003.

\bibitem{BorPapTso11}
L.~Borcea, G.~Papanicolaou, and C.~Tsogka.
\newblock Adaptive time-frequency detection and filtering for imaging in heavy
  clutter.
\newblock {\em SIAM J. Imaging Sciences}, 4(3):827--849, 2011.

\bibitem{BraWer65}
H.~Brakhage and P.~Werner.
\newblock \"{U}ber das {D}irichletsche {A}ussenraumproblem f\"ur die
  {H}elmholtzsche {S}chwingungsgleichung.
\newblock {\em Arch. Math.}, 16:325--329, 1965.

\bibitem{BurMinRam13}
C.~Burkard, A.~Minut, and K.~Ramdani.
\newblock Far field model for time reversal and application to selective
  focusing on small dielectric inhomogeneities.
\newblock {\em Inverse Problems and Imaging}, 7(2):445--470, 2013.

\bibitem{BurMil70}
A.~J. Burton and G.~F. Miller.
\newblock The application of integral equation methods to the numerical
  solution of some exterior boundary-value problems.
\newblock {\em Proc. Roy. Soc. London. Ser. A}, 323:201--210, 1971.
\newblock A discussion on numerical analysis of partial differential equations
  (1970).

\bibitem{CassierHazard}
M.~Cassier and C.~Hazard.
\newblock { Multiple scattering of acoustic waves by small sound-soft obstacles
  in two dimensions: Mathematical justification of the Foldy-Lax model}.
\newblock {\em {Wave Motion}}, {50}({18-28}), {2013}.

\bibitem{ChenLeeLin}
J.T. Chen, Y.T. Lee, Y.J. Lin, I.L. Chen, and J.W. Lee.
\newblock { Scattering of sound from point sources by multiple circular
  cylinders using addition theorem and superposition technique}.
\newblock {\em {Numerical Methods for Partial Differential Equations}},
  {27}({1365-1383}), {2011}.

\bibitem{ColKre98}
D.~Colton and R.~Kress.
\newblock {\em Inverse Acoustic and Electromagnetic Scattering Theory},
  volume~93 of {\em Applied Mathematical Sciences}.
\newblock Springer-Verlag, Berlin, second edition, 1998.

\bibitem{ColKre83}
D.~L. Colton and R.~Kress.
\newblock {\em Integral Equation Methods in Scattering Theory}.
\newblock Pure and Applied Mathematics (New York). John Wiley \& Sons Inc., New
  York, 1983.
\newblock A Wiley-Interscience Publication.

\bibitem{ISI:000259270600074}
A.~Devilez, B.~Stout, N.~Bonod, and E.~Popov.
\newblock {Spectral analysis of three-dimensional photonic jets}.
\newblock {\em {Optics Express}}, {16}({18}):{14200--14212}, {2008}.

\bibitem{ISI:000249155100055}
T.E. Doyle, D.A. Robinson, S.B. Jones, K.H. Warnick, and B.L. Carruth.
\newblock {Modeling the permittivity of two-phase media containing monodisperse
  spheres: Effects of microstructure and multiple scattering}.
\newblock {\em {Physical Review B}}, {76}({5}), {2007}.

\bibitem{ISI:000263911800054}
T.E. Doyle, A.T. Tew, K.H. Warnick, and B.L. Carruth.
\newblock {Simulation of elastic wave scattering in cells and tissues at the
  microscopic level}.
\newblock {\em {Journal of the Acoustical Society of America}},
  {125}({3}):{1751--1767}, {2009}.

\bibitem{MEhrhardtBook}
M.~Ehrhardt.
\newblock {\em Wave Propagation in Periodic Media Analysis, Numerical
  Techniques and practical Applications, E-Book Series Progress in
  Computational Physics (PiCP), Volume 1}.
\newblock Bentham Science Publishers, 2010.

\bibitem{MHZ}
M.~Ehrhardt, H.~Han, and C.~Zheng.
\newblock {Numerical simulation of waves in periodic structures}.
\newblock {\em {Commun. Comput. Phys.}}, {5}:{849--870}, {2009}.

\bibitem{ISI:000256469800019}
P.~Ferrand, J.~Wenger, A.~Devilez, M.~Pianta, B.~Stout, N.~Bonod, E.~Popov, and
  H.~Rigneault.
\newblock {Direct imaging of photonic nanojets}.
\newblock {\em {Optics Express}}, {16}({10}):{6930--6940}, {2008}.

\bibitem{Fin06}
M.~Fink.
\newblock Time-reversal acoustics.
\newblock In {\em Inverse problems, multi-scale analysis and effective medium
  theory}, volume 408 of {\em Contemp. Math.}, pages 151--179. Amer. Math.
  Soc., Providence, RI, 2006.

\bibitem{fin08}
M.~Fink.
\newblock Time-reversal acoustics.
\newblock {\em J. Phys.: Conf. Ser.}, 118(1):012001, 2008.

\bibitem{FinPra94}
M.~Fink and C.~Prada.
\newblock Eigenmodes of the time-reversal operator: A solution to selective
  focusing in multiple-target media.
\newblock {\em Wave Motion}, 20:151--163, 1994.

\bibitem{GeuzaineBrunoReitich}
C.~Geuzaine, O.~Bruno, and F.~Reitich.
\newblock {On the O(1) solution of multiple-scattering problems}.
\newblock {\em {IEEE Trans. Magn.}}, {41}({5}):{1488--1491}, {May} {2005}.
\newblock {11th IEEE Biennial Conference on Electromagnetic Field Computation,
  Seoul, South Korea, June 06-09, 2004}.

\bibitem{Greengard-Rokhlin}
L.~Greengard and V.~Rokhlin.
\newblock A fast algorithm for particle simulations.
\newblock {\em J. Comput. Phys.}, 73(2):325--348, 1987.

\bibitem{Grote2004630}
M.J. Grote and C.~Kirsch.
\newblock Dirichlet-to-{N}eumann boundary conditions for multiple scattering
  problems.
\newblock {\em J. Comput. Phys.}, 201(2):630 -- 650, 2004.

\bibitem{HarringtonMautz}
R.F. Harrington and J.R. Mautz.
\newblock {H}-field, {E}-field and combined field solution for conducting
  bodies of revolution.
\newblock {\em Archiv Elektronik und Uebertragungstechnik}, 4(32):157--164,
  1978.

\bibitem{HazRam04}
C.~Hazard and K.~Ramdani.
\newblock Selective acoustic focusing using time-harmonic reversal mirrors.
\newblock {\em SIAM J. Appl. Math.}, 64(3):1057--1076, 2004.

\bibitem{ISI:000253764200069}
P.~Hewageegana and V.~Apalkov.
\newblock {Second harmonic generation in disordered media: Random resonators}.
\newblock {\em {Physical Review B}}, {77}({7}), {2008}.

\bibitem{HuLu}
Z.~Hu and Y.Y. Lu.
\newblock Compact wavelength demultiplexer via photonic crystal multimode
  resonators.
\newblock {\em J. Opt. Soc. Amer. B}, to appear 2014.

\bibitem{Joan}
R.D.~Meade J.D.~Joannopoulos and J.N. Winn.
\newblock {\em Photonic Crystals: Molding the Flow of Light}.
\newblock Princeton University Press, 1995.

\bibitem{KharlamovFilip}
A.A. Kharlamov and P.~Filip.
\newblock { Generalisation of the method of images for the calculation of
  inviscid potential flow past several arbitrarily moving parallel circular
  cylinders}.
\newblock {\em {Journal of Engineering Mathematics}}, {77}({1}), {2012}.

\bibitem{Mar06}
P.~A. Martin.
\newblock {\em Multiple Scattering. Interaction of Time-Harmonic Waves with $N$
  Obstacles}, volume 107 of {\em Encyclopedia of Mathematics and its
  Applications}.
\newblock Cambridge University Press, Cambridge, 2006.

\bibitem{MasNacWaa97}
T.D. Mast, A.I. Nachman, and R.C. Waag.
\newblock Focusing and imaging using the eigenfunctions of the scattering
  operator.
\newblock {\em J. Acoust. Soc. Am.}, 102:715--725, 1997.

\bibitem{ISI:000249786400042}
H.~Mertens, A.~F. Koenderink, and A.~Polman.
\newblock {Plasmon-enhanced luminescence near noble-metal nanospheres:
  Comparison of exact theory and an improved Gersten and Nitzan model}.
\newblock {\em {Physical Review B}}, {76}({11}), {2007}.

\bibitem{Natarov}
D.M. Natarov, V.O. Byelobrov, R.~Sauleau, T.M. Benson, and A.I. Nosich.
\newblock { Periodicity-induced effects in the scattering and absorption of
  light by infinite and finite gratings of circular silver nanowires}.
\newblock {\em {Optics Express}}, {19}({22176-22190}), {2011}.

\bibitem{Ned01}
J.-C. N{\'e}d{\'e}lec.
\newblock {\em Acoustic and Electromagnetic Equations. Integral Representations
  for Harmonic Problems}, volume 144 of {\em Applied Mathematical Sciences}.
\newblock Springer-Verlag, New York, 2001.

\bibitem{PashaevYilmaz}
O.K. Pashaev and O.~Yilmaz.
\newblock {Power-series solution for the two-dimensional inviscid flow with a
  vortex and multiple cylinders}.
\newblock {\em {Journal of Engineering Mathematics}}, {65}({2}), {2009}.

\bibitem{Pra97}
C.~Prada.
\newblock The {D.O.R.T.} method.
\newblock {\em J. Acoust. Soc. Am.}, 101(5):3090--3090, 1997.

\bibitem{Saad}
Y.~Saad.
\newblock {\em Iterative Methods for Sparse Linear Systems}.
\newblock PWS Publishing Company Boston, 1996.

\bibitem{PhysRevA.90.023839}
R.~Savo, M.~Burresi, T.~Svensson, K.~Vynck, and D.S. Wiersma.
\newblock Walk dimension for light in complex disordered media.
\newblock {\em Phys. Rev. A}, 90:023839, Aug 2014.

\bibitem{ThierryThesis}
B.~Thierry.
\newblock {\em Analyse et Simulations Num\'eriques du Retournement Temporel et
  de la Diffraction Multiple}.
\newblock Nancy University, Th\`ese de Doctorat, 2011.

\bibitem{Thi14}
B.~Thierry.
\newblock A remark on the single scattering preconditioner applied to boundary
  integral equations.
\newblock {\em Journal of Mathematical Analysis and Applications}, 413(1):212
  -- 228, 2014.

\bibitem{ThoWuFin96}
J.-L. Thomas, F.~Wu, and M.~Fink.
\newblock Time reversal focusing applied to lithotripsy.
\newblock {\em Ultrasonic Imaging}, 18(2):106--121, 1996.

\bibitem{Tsang}
L.~Tsang, J.A. Kong, K.H. Ding, and C.O. Ao.
\newblock {\em Scattering of Electromagnetic Waves, Numerical Simulation}.
\newblock Wiley Series in Remote Sensing. J.A. Kong, Series Editor, 2001.

\bibitem{TuluYilmaz}
S.~Tulu and O.~Yilmaz.
\newblock {Motion of vortices outside a cylinder}.
\newblock {\em {Chaos}}, {20}({4}), {2010}.

\bibitem{Van}
B.~Van~Genechten, B.~Bergen, D.~Vandepitte, and W.~Desmet.
\newblock { A Trefftz-based numerical modelling framework for Helmholtz
  problems with complex multiple-scatterer configurations}.
\newblock {\em {J. Comput. Phys.}}, {229}({6623-6643}), {2010}.

\bibitem{NaturePhotonics}
D.S. Wiersma.
\newblock Disordered photonics.
\newblock {\em Nature Photonics}, 7:188--196, Feb 2013.

\end{thebibliography}

\end{document}